# Rethinking OWL Expressivity: Semantic Units for FAIR and Cognitively Interoperable Knowledge Graphs

*Why OWLs Don't Have to Understand Everything They Say*

Vogt, Lars[1];

[1]*TIB Leibniz Information Centre for Science and Technology, Welfengarten 1B, 30167 Hanover, Germany,* orcid.org/0000-0002-8280-0487

Correspondence to: lars.m.vogt@googlemail.com

—

# Abstract

Semantic knowledge graphs are foundational to implementing the FAIR Principles, yet RDF/OWL representations often lack the semantic flexibility and cognitive interoperability required in scientific domains. We present a novel framework for semantic modularization based on *semantic units*—modular, semantically coherent subgraphs enhancing expressivity, reusability, and interpretability—combined with *four new representational resource types* (some-instance, most-instances, every-instance, all-instances) for modelling assertional, contingent, prototypical, and universal statements. The framework enables the integration of knowledge modelled using different logical frameworks (e.g., OWL, First-Order Logic, or none), provided each semantic unit is internally consistent and annotated with its logic base. This allows, for example, querying all OWL 2.0-compliant units for reasoning purposes while preserving the full graph for broader knowledge discovery. Our framework addresses twelve core limitations of OWL/RDF modeling—including negation, cardinality, complex class axioms, conditional and directive statements, and logical arguments—while improving cognitive accessibility for domain experts. We provide schemata and translation patterns to demonstrate semantic interoperability and reasoning potential, establishing a scalable foundation for constructing FAIR-aligned, semantically rich knowledge graphs.

**Keywords**: FAIR Principles, knowledge graph, semantic interoperability, cognitive interoperability, semantic unit, OWL, logical framework, semantic modularization

# Background

As the volume of data generated daily continues to grow rapidly (1–3), modern research data infrastructures are increasingly expected to adhere to the **FAIR Guiding Principles** (4), which support the automation of data management processes by machines. This calls for data structures and corresponding tooling and mechanisms that ensure knowledge and data are **F**indable, **A**ccessible, **I**nteroperable, and **R**eusable—by both machines and humans.

**Knowledge graphs**, **semantic graph schemata**, and **ontologies** are instrumental in establishing the technical and conceptual frameworks essential for generating FAIR data and metadata. As key tools for organizing, connecting, integrating, and representing diverse data sources, they promise a paradigm shift in how we access, navigate, comprehend, and derive insights from data. Accordingly, they have become cornerstones across a wide range of domains and contexts, including semantic search, deep reasoning, knowledge discovery, decision-making, natural language disambiguation, machine reading, and entity consolidation for Big Data and text analytics (5–7).

A critical step in constructing knowledge graphs is **semantic parsing**—the process of translating information from various data sources into structured representations based on formal semantics interpretable by machines. This involves transforming a **source model** (e.g., a relational database, CSV file, or natural language expressions) into a **target model** that adopts the *Subject—Predicate—Object* triple syntax of the Resource Description Framework (RDF). The target model is built using ontologies and a knowledge representation language such as the Web Ontology Language (OWL) that is grounded in Description Logics. Although both source and target model represent models of the **same referent system**, they differ semantically and structurally (8). The target model is specifically designed for tasks like **reasoning**, **information integration**, and **information discovery through semantic search**. The goal of semantic parsing is thus to create a target model that preserves the source model's meaning while enhancing its FAIRness and machine-actionability through standardized, ontology-based vocabularies and semantic graph schemata (8).

Semantic parsing can be performed by humans, machines (via natural language processing), or by hybrid approaches that combine both. Regardless of method, it presents significant challenges for the rapid generation of FAIR and high-quality knowledge graphs. Ambiguities, context-dependent meanings, and linguistic nuances often necessitate deep domain expertise and the skills of ontology engineers to accurately model information into triples. Misinterpretations and imperfect representations can arise, and when data are automatically extracted and integrated, errors can propagate through a knowledge graph—potentially distorting downstream research.

In recent years, **Large Language Models** (LLMs) have gained prominence in enhancing access to scientific knowledge (9–15). They also have been employed to support the generation of knowledge graphs and ontologies and have been integrated with them to improve the LLM's own commonsense reasoning capabilities (16–23). However, despite this groundbreaking progress, constructing knowledge graphs and ontologies with the semantic depth, precision, and structural sophistication required in scientific contexts still demands deep domain expertise and the active involvement of ontology engineers. Knowledge graphs and ontologies cannot yet be reliably generated by LLMs alone (24,25).

The lack of a unified semantic framework and standardized formalism for representing data and knowledge in a knowledge graph—both crucial for achieving true interoperability—further complicates collaboration and data sharing. Although a common semantic infrastructure is desirable, its realization is hindered by diverse standards, conflicting ontologies, differing purposes of the underlying models, and the inherent subjectivity of semantic interpretation. These challenges make truly seamless interoperability between knowledge graphs difficult, if not impossible, to achieve (8).

A fundamental obstacle lies in the absence of a consistent semantic framework and formalism. Different technical implementations of knowledge graphs exist, leading to ambiguity in the term itself (6). A knowledge graph might be managed using a **labeled property graph** (e.g., [Neo4J](Neo4J)) or a **triple store based on RDF**. Moreover, not all RDF graphs are grounded in logic-based formalisms like OWL is in Description Logics. While other logics such as First-Order Logic exist, they remain underused due to scalability and tooling issues. **OWL** offers a robust and widely adopted framework, supporting formal semantics and reasoning as a W3C standard, whereas no such recommendation yet exists for labeled property graphs like Neo4j, despite proposals such as [OWLStar](OWLStar) and [owl2lpg](owl2lpg). However, OWL is often perceived as overly complex and lacking in **cognitive interoperability** (for cognitive interoperability, see Box 1). This perception deters non-experts in formal semantics and logic from effectively using OWL-based graphs. Conversely, labeled property graphs prioritize simplicity and intuitive structure but lack the formal semantics necessary for rigorous scientific applications. Thus, each framework presents trade-offs between formal expressivity and cognitive interoperability.

> **Box 1 | Cognitive Interoperability** (25)
>
> Cognitive interoperability is "a critical characteristic of data structures and information technology systems that plays an essential role in facilitating efficient communication of data and metadata with human users. By providing intuitive tools and functions, systems that support cognitive interoperability enable users to gain an overview of data, locate data they are interested in, and explore related data points in semantically meaningful and intuitive ways. The concept of cognitive interoperability encompasses not only how humans prefer to interact with technology, i.e., **human-computer interaction**, but also how they interact with information, i.e., **human information interaction**, considering their general cognitive conditions. In the context of information technology systems such as KGs, achieving cognitive interoperability necessitates tools that increase the user's awareness of the system's contents, that aid in understanding their meaning, support data and metadata communication, enhance content trustworthiness, facilitate integration into other workflows and software tools, and that clarify available actions and data operations. Additionally, cognitive interoperability also encompasses ease of implementation of data structures and their management for developers and operators of information technology systems. It thus addresses the specific data, tool, and service needs of the three main personas (26) identified for users of information management systems such as KGs, namely **information management system builders** (i.e., information architects, database admins), **data analysts** (i.e., researchers, data scientists, machine learning experts), and **data consumers** (i.e., stakeholders, end users, domain experts)" (p. 11-12) (25). Information technology systems and data and knowledge structures that are characterized by a high level of cognitive interoperability enable low-threshold access for their human users.

Balancing cognitive interoperability with formal expressivity is a persistent challenge in the development of knowledge graphs. OWL, rooted in Description Logics, imposes strict constraints on expressivity, while Neo4J provides no formal semantics.

The *Methods* chapter introduces a representational framework grounded in the concept of **semantic units** (27), along with **four novel representational resource categories**: some-instance, most-instances, every-instance, and all-instances resource. This framework implements the concept of **semantic modularization**—structuring the knowledge graph into semantically meaningful, independently addressable units that can be internally consistent with different logical frameworks.

In the *Challenges and their corresponding semantic unit approaches* chapter, we discuss twelve modelling and usability challenges that OWL-based and labeled-property-graph-based knowledge graphs face, each followed by a potential solution based on our representational framework. Since the challenges arise from OWL's underlying logic, they are difficult to resolve within the Description Logics framework or can only be represented in ways that limit the cognitive interoperability of the resulting representations. Nonetheless, our potential solutions demonstrate that alternative formalisms for representing specific types of statements in OWL are possible, provided that the representations are not intended to support automated reasoning. While not all proposed representations enable inferencing, they facilitate the systematic representation and classification of otherwise problematic types of information, thereby improving their discoverability and cognitive interoperability. Our approach thus shifts the balance of the twelve identified challenges toward greater expressivity and cognitive interoperability. This approach also aligns with the broader principle that, while machine-interpretability of all data and knowledge is desirable, it is more critical to ensure that all relevant content is represented. In this sense, *OWLs do not always have to understand everything they say—as long as they deliver the message*. By enabling clear categorization and explicit differentiation of these statements, we enhance both their semantic and cognitive interoperability within knowledge graphs.

In the subsequent *Discussion* chapter, we provide an overview of the taxonomy of proposed semantic unit types and discuss the granularity of semantic units and the logical reasoning possibilities in OWL-extended knowledge graphs. We also discuss the general technology-agnostic applicability of semantic units and provide key criteria that have to be implemented in knowledge management systems for applying the here proposed semantic units approach.

> **Box 2 | Conventions**
>
> In this paper, we refer to FAIR knowledge graphs as machine-actionable semantic graphs for documenting, organizing, and representing lexical, assertional (e.g., empirical data), contingent, prototypical, interrogation, directive, conditional, and universal statements, thereby contrasting knowledge graphs with ontologies, which contain mainly lexical and universal statements. We want to point out that we discuss semantic units against the background of RDF-based triple stores, OWL, and Description Logics as a formal framework for inferencing, and labeled property graphs as an alternative to triple stores, because these are the main technologies and logical frameworks used in knowledge graphs that are supported by a broad community of users and developers and for which accepted standards and tools exist. We are aware of the fact that alternative technologies and frameworks exist that support an n-tuples syntax and more advanced logics (e.g., First-Order Logic) (28,29), but supporting tools and applications are missing or are not widely used to turn them into well-supported, scalable, and easily usable knowledge graph applications.
>
> Throughout this text, we use <u>regular underlined</u> to indicate ontology classes, *<u>italicsUnderlined</u>* when referring to properties (i.e., relations in Neo4j), and use ID numbers to specify each. ID numbers are composed of the ontology prefix followed by a colon and a number, e.g., *<u>hasQuality</u>* (RO:0000086). If the term is not yet covered in any ontology, we indicate it with an asterisk (*). Newly introduced classes and properties relating to semantic units have the ontology prefix SEMUNIT as in the class *<u>SEMUNIT:has semantic unit subject</u>**. They will be part of a future Semantic Unit ontology. We use <u>'regular underlined'</u> to indicate instances of classes, with the label typically referring to the class label and the ID number to the class.
>
> When we use the term *resource*, we understand it to be something that is uniquely designated (e.g., a Uniform Resource Identifier, URI) and about which you want to say something. It thus stands for something and represents something you want to talk about. In RDF, the *Subject* and the *Predicate* in a triple are always resources, whereas the *Object* can be either a resource or a literal. Resources can be either properties, instances, or classes, with properties taking the *Predicate* position in a triple and with instances referring to individuals (=particulars) and classes to universals or kinds.
>
> For reasons of clarity, in the text and in all figures, we represent resources not with their URIs but with human-readable labels, with the implicit assumption that every property, every instance, and every class have their own URI. Additionally, we use the term *triple* to refer specifically to a triple statement, while the term *statement* pertains to a natural language statement, establishing a clear distinction between the two.

# Methods

To address the challenges outlined in the following chapter, we propose a representational framework based on two interrelated concepts: **semantic units** and a novel representational convention involving four new categories of **representational entities**. Together, these concepts serve as foundational building blocks for enhancing the expressiveness and cognitive interoperability of data and knowledge in knowledge graphs—particularly in scientific contexts where precision, complexity, and provenance are critical.

Semantic units implement **semantic modularization** by structuring a knowledge graph into modular, identifiable components that represent meaningful scientific content—such as assertions, procedures, or arguments. Semantic modularization ensures that each semantic unit maintains internal semantic coherence and, where applicable, adherence to a specific logical framework such as Description Logics (via OWL) or First-Order Logic, allowing for heterogeneous modelling, modular reasoning, and improved findability, accessibility, and reuse. Semantic units are both reusable and annotatable, enabling complex representations to be constructed from simpler parts. Complementing this modularization, we introduce four new categories of representational resources: **some-instance**, **most-instances**, **every-instance**, and **all-instances resources**. These categories extend the standard RDF/OWL resource model beyond named-individuals, classes, and properties, allowing for more faithful modelling of scientific statements central to scientific discourse and reasoning that are underrepresented in existing semantic frameworks.

By integrating these two concepts—semantic units and new resource categories—we provide a practical approach to addressing the twelve challenges identified in the following chapter. This integration supports the development of more semantically rich and cognitively interoperable knowledge graphs.

## Semantic units as modular building blocks for knowledge graphs

Semantic units organize a knowledge graph into **semantically meaningful subgraphs**—sets of tripes that represent individual **units of representation** (27). Unlike conventional graph partitioning approaches, which focus on technical or structural criteria (30–38), our method structures graphs from the perspective of domain experts, prioritizing **semantic coherence and cognitive interoperability**. This enables **semantic modularization**: the decomposition of a knowledge graph into logically coherent, semantically modelled units of information that can be independently reused, reasoned over, or queried.

A semantic unit is modeled as a **Named Graph**[1], identified by a **Globally Unique Persistent and Resolvable Identifier** (**GUPRI**). This dual use—serving as both the graph's URI and its referential anchor—allows users to treat the semantic unit both as an entity and the data it represents. For instance, the assertion "*Apple X has a weight of 204.56g*" can be encoded as a semantic unit with its own GUPRI, making it traceable, reusable, and annotatable independently of the surrounding graph, thus facilitating making statements *about* this statement—or, in general, statements about the content of any given semantic unit (27). Furthermore, the semantic unit resource instantiates a corresponding **semantic unit class**, thereby classifying its content.

---

[1] For a labeled property graph, by using corresponding key-value pairs that indicate the semantic unit's GUPRI for each relation belonging to the same semantic unit.

This structuring of a knowledge graph introduces an additional layer of triples to the graph. We therefore distinguish between:

- The **data graph layer**, which contains the domain-specific content in the form of RDF triples (e.g., measurement data, class assertions).
- The **semantic-units graph layer**, which holds metadata (e.g., who asserted the statement, under what schema) and organizational relations about semantic units (e.g., associated semantic units in a nested semantic units structure), and thus triples that have been added when introducing semantic units to the graph.

This layering ensures that both the information content and its contextual metadata are independently addressable. The distinction is not only applied to the graph as a whole, but also to every semantic unit (see Fig. 1C). Each semantic unit consists of:

- A **data graph** (its content),
- A **semantic-units graph** (its metadata and organizational links).

All triples involving the semantic unit resource itself reside in the semantic-units graph layer. This architecture enables straightforward specification of **metadata** such as creator, creation date, contributors, last updated on, copyright license, and versioning (see Fig 1C; the various metadata properties are indicated by *some metadata property* as a placeholder).

Semantic units are categorized into two top-level categories: statement units, which represent individual statements, and compound units, which aggregate multiple semantic units into single coherent representational structures.

**Statement unit**

A **statement unit** captures the **smallest unit of propositional meaning** in a knowledge graph. Each triple in the graph belongs to exactly one statement unit, forming a **mathematical partition** of the overall graph. The data graph layer of a knowledge graph can thus be viewed as the sum of all its statement unit data graphs. Every statement unit possesses a **subject resource** and one or more object resources or literals. The number of RDF triples in a statement unit and with it the number of object resources or literals varies depending on the **arity** of the underlying predicate or verb (Fig. 1C).

The data graph of each statement unit follows a corresponding **semantic graph schema**—a reusable template (e.g., SHACL shapes (39) or [LinkML](#) models) for representing a specific type of proposition in RDF/OWL—to ensure consistency, facilitate validation, and support propositional interoperability (8). These schemata can also support **user interface (UI) display patterns**, such as:

- **Dynamic labels**: human-readable, text-based representations for use in HTML or documentation, using labels and values from *Subject* and *Object* nodes of the corresponding statement units that carry information relevant to a human user (see Fig. 1D).
- **Dynamic mind maps**: mind map like graphical displays that map the same *Subject* and *Object* nodes to corresponding nodes in graphical representations (see Fig. 1E).

To enable referencing statements within statements—such as quoting, asserting, specifying a relationship between two statements, or qualifying another statement—we define **complex statement units**. These units blend the features of statement and compound units: they contain their own data graph and refer to other statement units using the property *<u>SEMUNIT: has associated semantic unit</u>*  (Fig. 2). Complex statement units are modelled as a subclass of statement units.

### A) Statement

*Apple X has a weight of 204.56 grams*

### B) OWL/RDF Graph

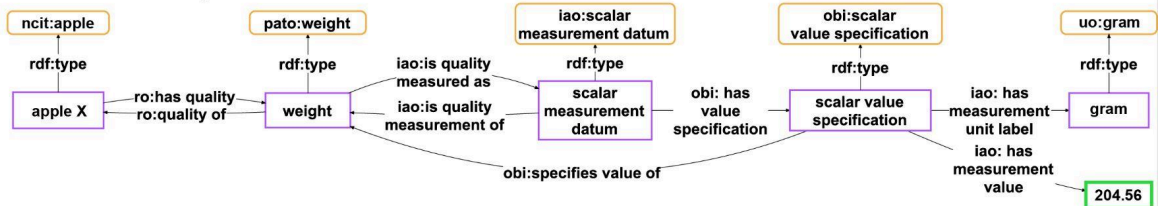

### C) Statement Unit

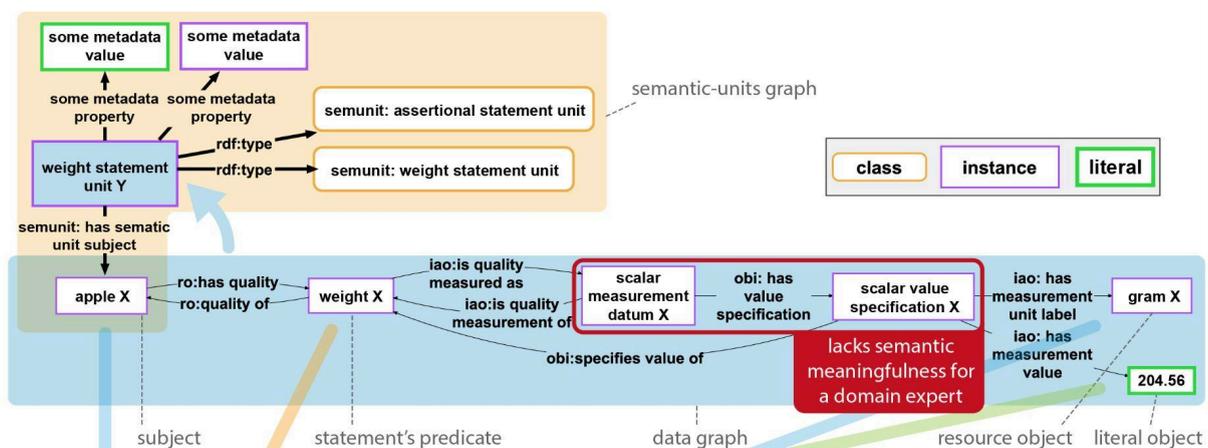

### D) Dynamic Label (Textual Display)

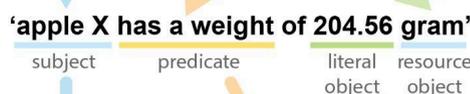

### E) Dynamic Mind Map (Graphical Display)

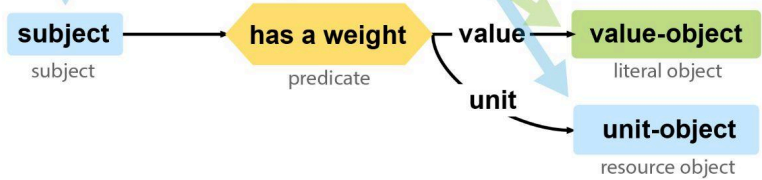

**Figure 1: From statement to graph to semantic unit. A)** A human-readable assertional statement about the weight measurement of an apple. **B)** The corresponding representation of the same statement as an instance-based semantic graph, adhering to RDF syntax and following the established pattern for measurement data from the Ontology of Biomedical Investigations (OBI) (40). **C)** The same graph, organized as a statement unit, which is a specific category of semantic unit. The data graph, denoted within the blue box at the bottom, articulates the statement with 'apple X' as its subject and 'gram X' alongside the numerical value 204.56 as its objects. Note, how the graph from B) forms the data graph of this statement unit. The peach-colored box encompasses its semantic-units graph. It explicitly denotes the resource embodying the statement unit (bordered blue box) as an instance of the *<u>SEMUNIT:weight statement unit</u>* class and the *<u>SEMUNIT:assertional statement unit</u>* class, with 'apple X' identified as the subject. Notably, the GUPRI of the statement unit is also the GUPRI of the semantic unit's data graph (the subgraph in the not bordered blue box). The semantic-units graph also contains various metadata triples, here only indicated by *some metadata property* and *some metadata value* as their placeholders. Highlighted in red within the data graph is an example of a triple that is required for modelling purposes but lacks semantic meaningfulness for a domain expert. **D)** The dynamic label associated with the statement unit class (*SEMUNIT:weight statement unit*). **E)** The dynamic mind map associated with the statement unit class (*SEMUNIT:weight statement unit*).

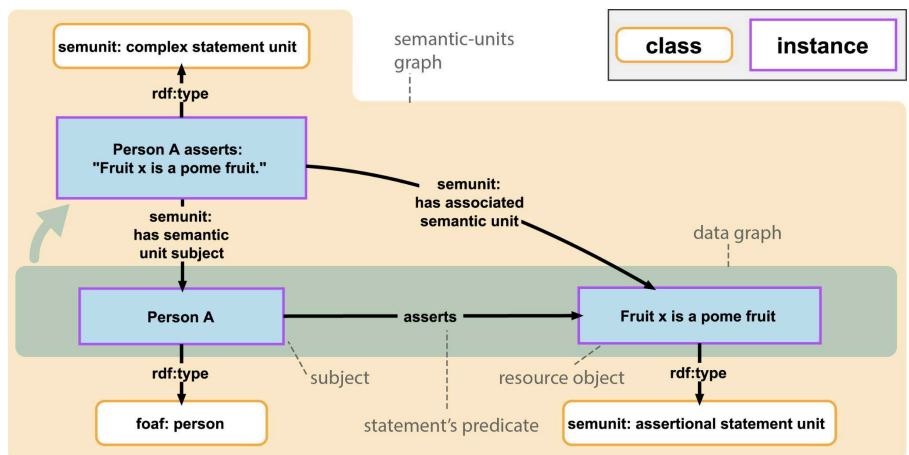

Figure 2: Example of a complex statement unit in which at least one semantic unit resource takes the subject or one of the object positions of the statement. Here, 'Person A' (FOAF:person) takes the subject position and the assertional statement unit resource *'Fruit X is a pome fruit'* the object position of a statement, forming the data graph of the complex statement unit.

**Statement unit metadata**

Since only statement units carry semantic content, they are also responsible for the following extended metadata (drawing on (8,25)):

1) The GUPRI of the semantic graph schema (e.g., SHACL shape) used to structure the data graph of the statement unit (to support propositional interoperability).
2) The author(s) of the content (which is not necessarily also the creator of the semantic unit).
3) The formal logical framework applied (e.g., OWL DL, First-Order Logic), enabling appropriate reasoning.
4) Ideally, the degree of certainty associated with the statement (e.g., confidence level or provenance-based evidence levels).

This metadata supports reasoning, versioning, traceability, reuse, and semantic interoperability—crucial for the FAIRification of scientific knowledge graphs. Additionally, if applicable, the source for the content can be added (e.g., a literature reference, when extracted from a publication) as well as the extraction method (e.g., manual or using a specific LLM).

**Compound unit**

A compound unit serves as a higher-level organizational resource that groups together a **semantically meaningful collection of semantic units** (either statement or compound units). Each compound unit is represented by its own GUPRI and instantiates a corresponding compound unit class (Fig. 3). Unlike statement units, compound units **do not contain a data graph**; instead, their meaning is derived entirely from the units they aggregate. Thus, they contribute exclusively to the semantic-units graph layer of a knowledge graph, and not to its data graph layer.

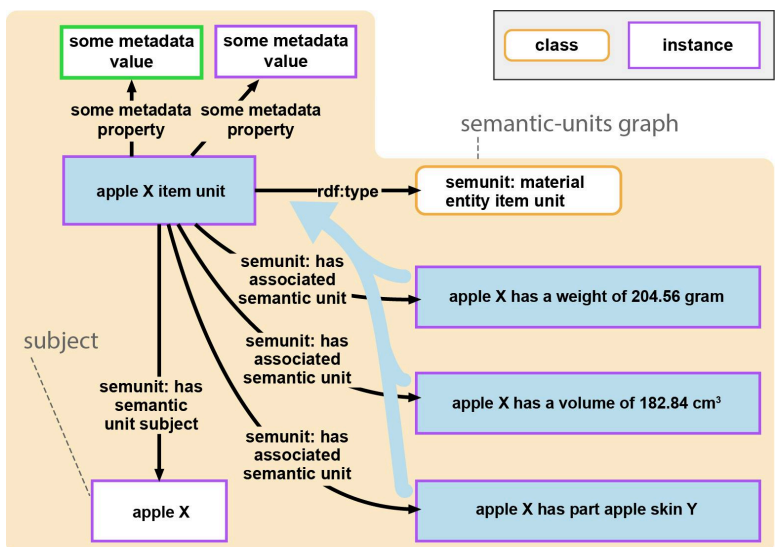

**Figure 3: Example of a compound unit**, denoted as *'apple X item unit'*, that encompasses multiple statement units describing apple X. Compound units, by virtue of merging the data graphs of their associated statement units, only indirectly manifest a data graph (here, highlighted by the blue arrow). Notably, the compound unit possesses a semantic-units graph (depicted in the peach-colored box) delineating the associated semantic units.

# A novel representational convention: some-instance, most-instances, every-instance, and all-instances resources

The OWL framework defines three fundamental categories of representational resources:

- **Named-individuals**: specific, known instances of one or more classes, representing particulars.
- **Classes**: abstract representations of kinds or universals.
- **Properties**: relationships between resources (object properties), between a resource and a literal (data properties), or for annotating resources (annotation properties).

While foundational, this classification is **insufficient for modelling universal**, **prototypical**, or **existential claims** common in scientific and empirical discourse. To address this limitation, we propose **four additional categories of representational resources** that extend OWL's taxonomy: some-instance, most-instances, every-instance, and all-instances resources.

### Some-instance resource

A **some-instance resource** denotes **one or more unspecified instances** of a target class $C$. It is used to express existential statements of the form: "*There exists at least one instance of class $C$ for which…*".

**Notation**: $\exists i \in C$, where $i$ denotes an instance and $C$ the class.

Formally, a some-instance resource is typed as both an instance of *SEMUNIT: some-instance resource* and an instance of the target class $C$.

When used as a subject or object in a triple, a some-instance resource implies that the triple applies to some non-empty subset of instances of $C$, with the triple intended to be interpreted as applying to each member of that subset. This characterization facilitates, for instance, the representation of portions of matter. To illustrate, a specific portion of water can be designated by associating an 'object aggregate' (BFO:0000027) with a some-instance resource of water (CHEBI:15377) through *hasPart* (BFO:0000051).

**Most-instances resource**

A **most-instances resource** refers to a **typical but not universal subset** of instances of class $C$. It allows modelling of statements like "*Most birds fly*", where a majority holds the property, but exceptions exist (e.g., ostriches, penguins).

Formally, this resource represents a subclass $D \subseteq C$, defined by distinguishing properties $A$. It fulfills the condition: $|D| > |C \setminus D|$; i.e., the cardinality of $D$ is greater than the cardinality of $C$ minus $D$. Each most-instances resource is typed as an instance of both *SEMUNIT: most-instances resource* and $C$.

The introduction of most-instances resources enables prototypical generalizations while avoiding incorrect universal claims.

**Every-instance resource**

An **every-instance resource** expresses **universal quantification** over a class $C$. It asserts that **a given property or relation holds for all members** of the class, individually.

**Notation**: $\forall i \in C$, where the semantics read: "*For every instance i of class C, it holds that…*"

The resource is declared as an instance of the class *SEMUNIT: every-instance resource* and an instance of $C$. This is useful for strict, logically universal claims such as: "*Every instance of water consists of hydrogen and oxygen atoms.*"

**All-instances resource**

An **all-instances resource** denotes the **entire set of instances** of a class $C$, across all time (past and present). It is used for **statistical** or **comparative statements** about a class's population. Possible use cases include comparing cardinalities (e.g., *"There are more prokaryotes than eukaryotes"*) and expressing distributional properties (e.g., average height, frequency). Formally, an all-instances resource is typed as both an instance of the class *SEMUNIT: all-instances resource* and the target class $C$.

In combination with semantic units, these additional resources enable the representation of complex scientific statements that traditional RDF/OWL cannot express directly. In the next chapter, we show how these constructs resolve twelve key challenges of modelling and accessing data and knowledge in FAIR knowledge graphs.

# Challenges and their corresponding semantic unit approaches

In the subsequent sections, we delineate twelve different challenges encountered in knowledge graph modelling and present corresponding semantic unit-based approaches as their potential solutions. Each section commences with a discussion of one or more challenges, followed by the semantic unit approach devised to address them. Notably, some approaches offer resolutions to multiple challenges; hence, some sections introduce more than one challenge before delving into a shared semantic unit strategy.

## Representing fundamental categories of statements in FAIR knowledge graphs

**Challenge 1: Unified representation of diverse statements categories**
Effective communication of information necessitates distinguishing between various statement types, each encapsulating different modalities of truth, generality, or reference. For knowledge graphs, particularly within scientific domains, it is imperative to represent these distinctions to ensure semantic and cognitive interoperability. We identify five fundamental categories of statements:

1) **Assertional statements**: These specify concrete **facts** about individual entities, reflecting **what is the case**—referring to the universe of actualities, the real. Assertional statements are propositions that are meant to be either *true* or *false* (41,42). For instance: *"By 27 February 2020, the outbreak of coronavirus disease 2019 (COVID-19) caused 82623 confirmed cases and 2858 deaths globally"* (p.568, (43)). In knowledge graphs, such statements link known instances (i.e., owl:namedIndividual) to each other, representing empirical data like observations and measurements.

2) **Contingent statements**: These indicate **what can be the case**—referring to the **universe of possibilities**. They represent propositions that are true for some (i.e., at least one), but not necessarily for every instance of a class (41,42,44). One can derive a contingent statement from a single assertional statement—if you saw a white swan, you know that swans can be white—a fact is a proof for a possibility. Sometimes, it does not make sense to make assertional statements about individual entities. In these cases, it is often more reasonable to talk about possibilities. For example: *"... SARS-CoV-2 … can cause acute respiratory distress syndromes (ARDS)"* (p.569 (43)). Such statements are prevalent in contexts where universal claims are untenable, like in genetics or biology in general.

3) **Prototypical statements**: A subset of contingent statements, these denote **what is typically but not necessarily the case** for the instances of a class. They reflect statistical tendencies or default expectations rather than absolute truths. For example: *"In the presence of native predators, morphological defences typically consist of developing deeper, longer, and more pigmented tails, ..."* (p. 8, (45)). Prototypical statements are common in most empirical fields, such as medicine and ecology, where exceptions are frequent (see also *cluster classes* and *fuzzy sets* in (46)).

4) **Universal statements**: These assert **what is necessarily or impossibly the case**—referring to the **universe of intentions or finality**. They encompass established domain knowledge, such as scientific laws, hypothesis, and definitions (see also *essentialistic classes* in (46)). For instance, the infamous example of a universal statement that was considered true by scholars in Europe until the 17th Century[2] (47): "...*all swans are white*" (p. 4, (48)).

5) **Lexical statements**: These pertain to **linguistic items**, discussing terms and their attributes as textual representational artifacts (cf. *terminological statement* sensu (41); see also Ingvar Johannson's distinction between *use* and *mention* of linguistic entities (49)). They include labels, synonyms, human-readable definitions, and provenance metadata, in ontologies often represented using annotation properties.

These fundamental categories of statements are central to scientific discourse. Accurately representing their formal semantics in knowledge graphs is challenging. Consider the example of the white swan applied across different statement categories (ignoring lexical statements for now):

1) **Assertional**: "*Swan Anton is white*", or "*This swan has quality this white*" (*this-to-this*[3] relation).
2) **Contingent**: "*Swans can be white*", or "*Some swan has quality some white*" (*some-to-some*[4] relations).
3) **Prototypical**: "*Swans are typically white*", or "*Most swans have quality some white*" (*most-to-some* relations).
4) **Universal**: "*All swans are white*", or "*Every swan has quality some white*" (*all-to-some* relations).

Modelling these distinctions requires more than just resources for '*swan*', '*white*', and '*has quality*'. In labeled property graphs, the lack of standardized semantics means such statements rely heavily on context for interpretation, leading to ambiguity.

OWL, grounded in Description Logics, differentiates between **universal statements** (terminology box; **TBox**; see Fig. 4C) and the **domain of discourse** of a knowledge graph in the form of **assertional statements** (assertion box; **ABox**; see Fig. 4B) (50). However, OWL lacks formal mechanisms to represent contingent and prototypical statements.

**A) Statement modeled in a labeled property graph**

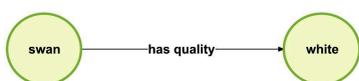

**B) OWL assertional statement mapped onto RDF**

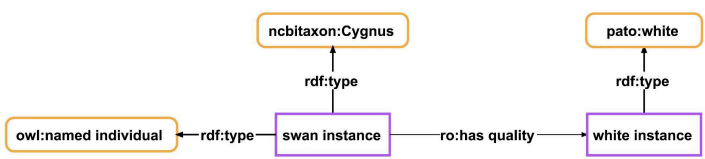

**C) OWL universal statement mapped onto RDF**

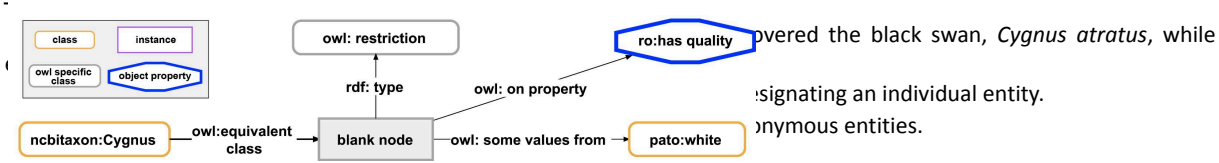

**Figure 4: Comparison of statement types as they can be documented in a labeled property graph and an OWL-based RDF graph. A)** A labeled property graph cannot formally distinguish assertional, contingent, prototypical, and universal statements and leaves it to the user to interpret the displayed relation. In OWL, mapped onto RDF, two different types of statements can [be] covered the black swan, *Cygnus atratus*, while [de]signating an individual entity. [an]onymous entities.

be distinguished: **B)** The assertional statement *This swan has quality this white* and **C)** *All swans have quality some white*. However, the latter cannot be documented in a knowledge graph but only be expressed as a class axiom within an ontology.

Whereas representing the universal statement "*All swans are white*" in OWL results in a complex but semantically precise representation of the statement (Fig. 4C), a domain expert will likely have problems understanding it and would prefer the simpler and still intuitively comprehensible representation provided by a labeled property graph (Fig. 4A). In other words, the OWL expression is lacking cognitive interoperability. Moreover, the OWL-based representation involves blank nodes, which makes basic SPARQL querying difficult (50) and has the potential to cause other issues (51) (see also *Challenge 3*).

While solutions exist for representing universal statements in labeled property graphs—such as annotating edges with logical properties (e.g., existential restriction axioms), resulting in an object-property mapping (50,52)—there is no standardized W3C recommendation for mapping OWL into property graphs.

In summary, the diversity of scientific statements underscores the limitations of current graph modelling formalisms. Labeled property graphs offer **cognitive interoperability** but lack **formal semantics**, whereas RDF/OWL-based graphs provide for some formal semantics but fall short in representing all statement types and fail in cognitive interoperability (for an overview of the differences between RDF/OWL-based knowledge graphs and labeled property graphs, see Table 1).

**Table 1: Comparison of the advantages and shortcomings of knowledge graphs based on RDF/OWL and on labeled property graphs** (50,52).

|  | **RDF/OWL** | **labeled property graph** |
| --- | --- | --- |
| query language | SPARQL, W3C recommendation | various, none has a W3C recommendation |
| formal semantics and reasoning | OWL; however:<br>1. when mapped to RDF, OWL is usually very verbose, leading to unnecessarily complex graphs that are not intuitively comprehensible for a human reader and inhibit effective computation;<br>2. OWL is not the only mode of inference, and other frameworks besides Description Logics exist for inferencing | no unified set of standards (no W3C recommendation)—a standardized mapping of OWL to a property graph is still lacking; the data model of Neo4j is not based on a formal set theory or First-Order Logic |
| relating information directly to edges | only indirectly through reification, RDF-star, or named graphs | information can be directly related to relations, i.e., edges |
| distinction of assertional, contingent, prototypical, and universal statements | formalized semantic distinction between assertional and universal statements; contingent and prototypical statements are not modelled in OWL | no formalized semantic distinction |

**Challenge 2: Representing universal statements in knowledge graphs**

In OWL, classes are identified by URIs, allowing explicit referencing. However, the axioms defining these classes lack URIs, rendering them unaddressable within the graph structure. Consequently, TBox axioms—despite being central to an ontology term's meaning—are not part of the domain of discourse in OWL-based knowledge graphs. A statement such as "*Author A contradicts class axiom X*" cannot be formally represented if *X* is only an implicit part of a class definition.

This limitation hinders the documentation of provenance, debates, or interpretations of universal statements in OWL-based knowledge graphs.

To address these challenges and enable comprehensive representation of assertional, contingent, prototypical, universal, and lexical statements, we introduce a semantic unit-based classification of statement types. This approach facilitates fine-grained representation and direct referencing of diverse knowledge forms. The subsequent section introduces this approach and outlines five key categories of semantic units designed to tackle these challenges.

**Approach: Distinguishing five categories of statement units**

To address the limitations outlined in *Challenges 1* and *2*, we introduce a semantic unit-based framework that enables the formal representation of diverse types of statements. In this framework, each individual statement in a knowledge graph is represented as a dedicated instance of a statement unit class (27) that is defined in a corresponding *Statement Unit Ontology*. These statement unit classes are organized along two orthogonal dimensions:

1) **Relation-based classification**: Groups statement units by the specific **ontological relation** they express (e.g., *has part*, *type*, *develops from*).
2) **Category-based classification**: Differentiates statement units by the nature of their **subject resource**, resulting in four mutually exclusive categories: assertional, contingent, prototypical, and universal statement units. We can further add the category of lexical statement units to this list, which comprise statements about linguistic items.

Because these classification criteria are independent, a single statement unit may simultaneously belong to a particular relation-based class and one of the five statement categories. For instance, the statement unit illustrated in Figure 1C is both a *weight statement unit* (relation-based) and an *assertional statement unit* (category-based). This dual categorization enables a flexible yet unified model for representing diverse statement categories within a knowledge graph.

Lexical statement unit

Lexical statement units capture metadata and linguistic annotations associated with the resources used in a knowledge graph. These include common annotation properties such as labels, synonyms, usage examples, elucidations, and editorial notes.

A key subclass of lexical statement units is the **identification unit**, which indicates the representational type of resource. We distinguish six forms of identification units, each corresponding to a specific resource type[5]:

- **Named-individual identification units** identify a resource as a named-individual resource of a particular target class and associate it with a label (Fig. 5A).
- **Some-instance**, **most-instances**, **every-instance**, and **all-instances identification units** each identify a resource as a corresponding type of resource of a particular target class, differing in scope or quantification (e.g., some, most, every, or all instances)(Fig. 5B-E).

---

[5] One could argue that the information about the class affiliation of the subject-resources does not belong to a lexical statement unit, as it relates to ontological properties of the resource. However, for pragmatic reasons concerning the organization and management of semantic units in knowledge graphs, we decided to include this information and call this type of lexical statement an identification unit.

- **Class identification units**[6] reference an ontology class and typically include metadata such as the class label, URI, the URI of the source ontology, and its ontology version (Fig. 5F).

Lexical statement units thus serve to annotate and clarify the representational nature and context of resources in a structured and machine-readable form.

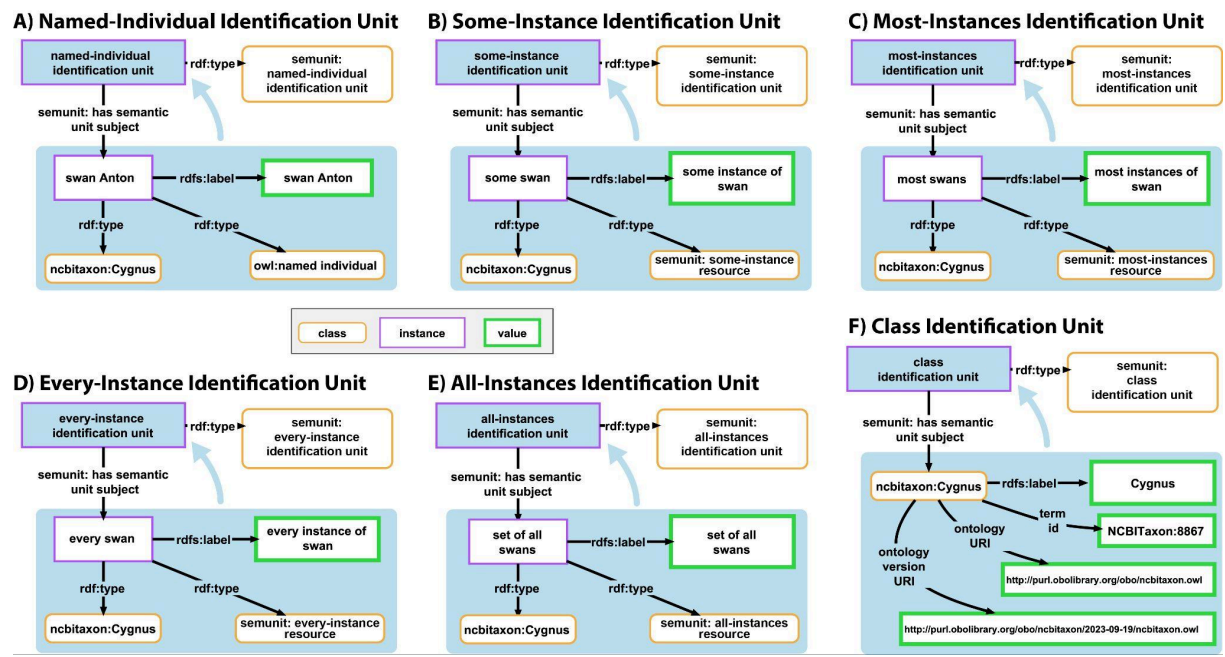

**Figure 5: Examples for six different types of identification units.** They are distinguished based on the type of subject resource they possess, which in turn are distinguished based on the property connecting that resource to its class resource. Each identification unit is represented by its own resource (blue box with borders), with its data graph shown by the blue box without borders. Except for the class identification unit (F), all subject resources of identification units are specified through the property *type* (RDF:type) as instances of two distinct classes, one of which is the type of representational resource and the other one the type of entity (here, the swan Cygnus (NCBITaxon:8867)). **A)** A **named-individual identification unit** that has the named-individual resource 'swan Anton' (NCBITaxon:8867) as its subject. The resource instantiates the class named individual (owl:namedIndividual) in addition to Cygnus (NCBITaxon:8867). **B)** A **some-instance identification unit** that has the some-instance resource 'some swan' (NCBITaxon:8867) as its subject, which instantiates the class SEMUNIT: some-instance resource in addition to Cygnus (NCBITaxon:8867). **C)** A **most-instances identification unit** that has the most-instances resource 'most swans' (NCBITaxon:8867) as its subject, which instantiates the class SEMUNIT: most-instances resource in addition to Cygnus (NCBITaxon:8867). **D)** An **every-instance identification unit** that has the every-instance resource 'every swan' (NCBITaxon:8867) as its subject and that instantiates the class SEMUNIT: every-instance resource in addition to Cygnus (NCBITaxon:8867). **E)** An **all-instances identification unit** that has the all-instances resource 'all swans' (NCBITaxon:8867) as its subject and that instantiates the class SEMUNIT: all-instances resource in addition to Cygnus (NCBITaxon:8867). **F)** A **class identification unit** that models the label and the identifier of the class Cygnus (NCBITaxon:8867). Optionally, it also provides the URI of the ontology and that of the ontology version from which the class has been taken. *For reasons of clarity, metadata for each semantic unit is not represented.*

Assertional statement unit

Assertional statement units express propositions about **named individuals**, i.e., explicitly identified entities within the knowledge graph. These statements correspond to classical ABox assertions in OWL, and are typed both as *SEMUNIT: assertional statement unit* (category-based) and as some relation-based class. When an object of the statement is another resource, it must also be a named individual. For example, Figure 6A shows a statement where the subject is the named-individual resource '<u>swan_Anton</u>' (NCBITaxon:8867). Assertional statement units are thus well-suited for expressing factual claims about known, context-specific entities.

---
[6] Analog to class identification units, one could additionally also specify property identification units that have property resources as their subject.

## Contingent statement unit

Contingent statement units make **existential claims** involving **anonymous instances** of a target class—referred to as **some-instance resource**. They capture propositions that are valid for at least one instance, enabling support for knowledge derived from empirical or partial evidence, representing observations or possibilities rather than universal truths. When a statement has a resource in its object position, it also takes the form of a some-instance resource (Fig. 6B). Each contingent statement unit is assigned both a category-based (*<u>SEMUNIT: contingent statement unit</u>*) and a relation-based type.

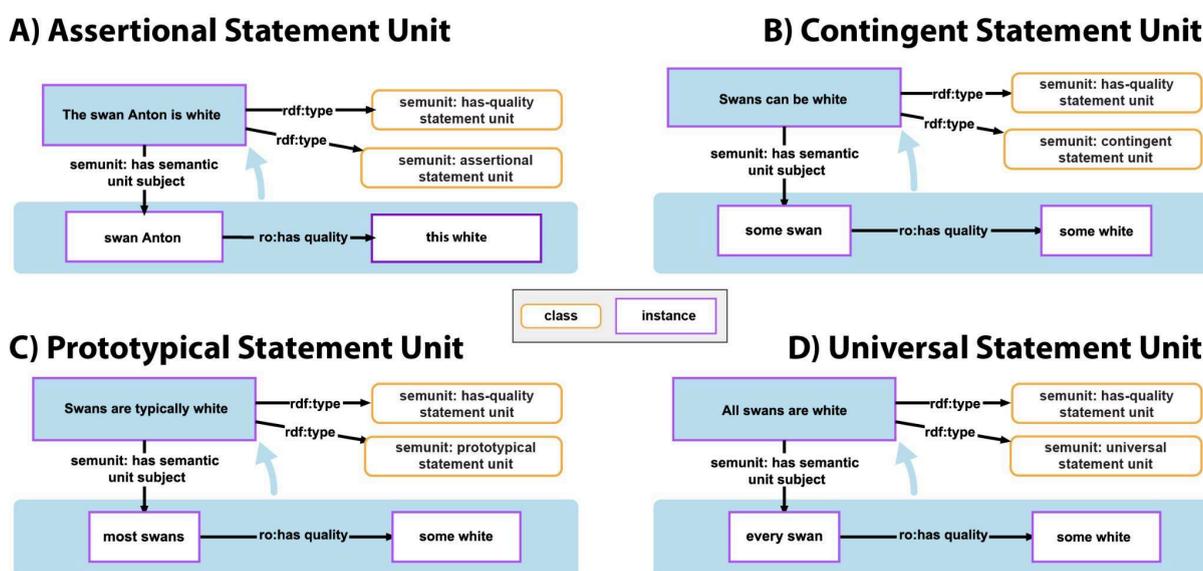

**Figure 6: Four subcategories of statement unit. A) Assertional statement unit.** This example models a has-quality relation between two named individuals. The data graph expressing this relation is shown in the blue box without borders. It is an instance-based data graph and forms an ABox. The subject of this assertional statement unit is an instance resource that is also the subject of a named-individual identification unit (cf. Fig. 5A). The assertional statement unit resource (blue box with borders, here shown with its dynamic label), is an instance of *<u>SEMUNIT: has-quality statement unit</u>* and of *<u>SEMUNIT: assertional statement unit</u>*. **B) Contingent statement unit.** This example models a has-quality relation that exists between some but not necessarily every instance of '<u>swan</u>' (NCBITaxon:8867) and some but not necessarily every instance of '<u>white</u>' (PATO:0000323). The subject of this contingent statement unit is also the subject of a some-instance identification unit (cf. Fig. 5B). The contingent statement unit resource (blue box with borders, here shown with its dynamic label), is an instance of *<u>SEMUNIT: has-quality statement unit</u>* and of *<u>SEMUNIT: contingent statement unit</u>*. **C) Prototypical statement unit.** This has-quality relation exists between most but not necessarily every instance of '<u>swan</u>' (NCBITaxon:8867) and some but not necessarily every instance of '<u>white</u>' (PATO:0000323). The subject of this prototypical statement unit is also the subject of a most-instances identification unit (cf. Fig. 5C). The prototypical statement unit resource instantiates the class *<u>SEMUNIT: prototypical statement unit</u>*. **D) Universal statement unit.** This example models a has-quality relation that exists between every instance of '<u>swan</u>' (NCBITaxon:8867) and some, but not necessarily every, instance of '<u>white</u>' (PATO:0000323). The data graph expressing this relation is shown in the blue box without borders and can, in principle, be translated into OWL to form a TBox. The subject of this universal statement unit is also the subject of an every-instance identification unit (cf. Fig. 5D). The universal statement unit resource (blue box with borders, here shown with its dynamic label), is an instance of *<u>SEMUNIT: universal statement unit</u>*. *For reasons of clarity, metadata for each semantic unit is not represented.*

Prototypical statement unit

Prototypical statement units represent **typical but not universal** characteristics of instances within a target class $C$. They use a **most-instances** resource as the subject, allowing the encoding of **prototypical relationships** and thus general patterns, tendencies, and empirical regularities, such as relationships between diseases and their symptoms or drugs and their effects, which are often expressed in reference to some probability evaluation and which are subject to exceptions (see also Fig. 6C).

**OWL and Description Logics do not natively support the notion of prototypicality**. To enable formal semantics for such statements (42), we define a subclass $D$ of the target class $C$, where every instance of $D$ possesses a property set $A$. A prototypical statement is then valid if the cardinality of $D$ exceeds that of the complement set ($C \setminus D$). This allows the expression of statements such as "*Most apples are green*", meaning that the number of green apples in class apple exceeds the number of non-green ones.

This approach opens the door for modelling prototypical statements within a structured framework.

Universal statement unit

Universal statement units express **generalizations that hold for all instances** of a given class. The subject is either the **class** itself or a quantified universal resource (**every-instance** or **all-instances**). Two structural patterns apply:

- **Class-based universal statements**: Use OWL constructs such as *subClassOf* (RDFS:subClassOf), *equivalentClass* (owl:equivalentClass), *disjointWith* (owl:disjointWith), or *sameAs* (owl:sameAs) to describe relationships between classes[7].
- **Quantified universal statements**: Have an every-instance or all-instances resource as their subject. If these statements include an object resource, it is typically a some-instance resource, resulting in an *all-to-some* relation. In n-ary statements, multiple object resources may be involved.

For example, the statement "Every swan (NCBITaxon:8867) *hasQuality* (RO:0000086) some white (PATO:0000323)," is represented as a universal statement unit with an every-instance subject (Fig. 6D). Universal statement units thus support the formalization of domain axioms and universally valid rules as instance-based graphs within a knowledge graph.

**Statement unit categories and formal semantics**

As discussed in previous sections, OWL's native semantics primarily support assertional and universal statements, leaving contingent and prototypical statements without a formal framework (see *Challenge 1*). The semantic units approach extends OWL's expressivity to encompass these additional statements types while maintaining compatibility with existing ontology infrastructure and while enhancing both their **cognitive interoperability** and **query properties** by lacking blank nodes and lowering the modelling complexity.

---

[7] If one understands OWL properties to refer to universals, the subject-resource could also be an OWL property and the statement based on corresponding properties such as *subPropertyOf* (owl:subPropertyOf), *domain* (owl:domain), or *range* (owl:range).

Semantic units do more than categorize knowledge—they make statements **first-class citizens** in the knowledge graph, and they **operationalize semantic modularization** by allowing each unit to be structured as a self-contained, semantically meaningful module that can be individually annotated, queried, reasoned over, or translated. This allows individual universal statements and consequently particular class axioms to be referenced and annotated directly, making them accessible to the general domain of discourse of a knowledge graph (see *Challenge 2*).

From a formal perspective, statement units can be treated as translations between ABox and TBox statements or as translations between logic programs and OWL axioms, as we previously outlined (53). Consider the assertional statement "'Swan Anton' (NCBITaxon:8867) *has-quality* (RO:0000086) 'white' (PATO:0000323)." This is modelled as a *has-quality* statement unit and corresponds to an ABox representation. Semantically, it can be translated into a TBox axiom using an ontology design pattern that depends on the type of statement unit. These translations can be formalized using expressions of **relational ontology design patterns** (54), which are OWL axioms involving variables for arbitrary entities. For example, the *'has-quality assertional statement unit'* translates to

> *?X SubClassOf: has-quality some ?Y,*

where *?X* and *?Y* correspond to the subject and object of the statement unit, respectively. These patterns can be represented as literals in the graph—using datatype or annotation properties—linked to the corresponding statement unit class. This allows OWL reasoners or SPARQL queries to retrieve all statements aligned with a given design pattern.

As discussed in (53), an alternative is to decouple the statement and its formal interpretation entirely, enabling a dual formalism. One part represents the semantic unit as a structured ABox statement; the other encodes its interpretation as a TBox axiom (or logic rule). This opens the door for **dual kind of reasoning** strategies: **OWL reasoning** for classical inference and consistency checking using the TBox axioms, and **logic programming** for more flexible, rule-based reasoning directly on the ABox statements.

The flexibility is crucial for representing **non-monotonic statements**—such as exceptions and defaults and thus contingent and prototypical statements—which cannot be expressed within OWL's strictly monotonic logic (55,56). For example, the prototypical statement "*Swans are typically white*" can be encoded using **Answer Set Programming (ASP)** (53) as:

> *has-quality(x, White) :- rdf:type(x, Swan), not lacks-quality(x, White).*

Here, *not* represents weak negation, reflecting the idea that the absence of evidence to the contrary implies the default. Under ASP semantics, such rules support **non-monotonic inference**: adding the fact *lacks-quality(x, White)* invalidates the previous default inference (*has-quality(x, White)*). This stands in contrast to OWL, where added facts cannot retract previous conclusions.

To bridge logic programs and OWL, we treat all OWL entities—classes, individuals, and properties—as individuals within the logic program. Each rule corresponds to an ontology design pattern and is translated into a set of OWL axioms. Grounders like Clingo (57) and solvers like DLV (58) or clasp (59) allow fully-grounded models to be computed. These models can then be forward-translated into OWL axioms (53).

An inverse translation is also feasible. For any given ontology design pattern, one can use an OWL reasoner to check whether the corresponding pattern is entailed by the ontology $O$'s deductive closure (O⁺). However, this requires querying over combinations of entities: for a pattern with n variables and |E| entities in the OWL ontology, this results in a computational complexity of $O(|E|n)$.

Each statement unit thus comprises two components:

1. A **logic program** that captures the instance-level representation.
2. A **pattern** that defines how predicates translate into OWL axioms.

For example, a named-individual identification unit (as in Figure 5A) is defined as:

Logic program:     *rdf:type(SawnAnton, ncbitaxon:Cygnus)*
Pattern:           *rdf:type(SawnAnton, ncbitaxon:Cygnus)*

Generalized, a named-individual identification unit could be represented as:

*named-individual-identification-unit(x), has-semantic-subject(x,y), rdf:type(y, z), rdfs:label(y, l), owl:named_individual(y)*;

with an appropriate OWL pattern, translating each predicate into its corresponding OWL statement.

Some-instance identification units follow the same pattern but do not assign a name to the instance, instead implicitly introducing it via "Skolemization". By contrast, every-instance identification units require a more involved modelling strategy. For example, *everySwan* is modelled as a collection whose members are all and only instances of Cygnus (NCBITaxon:8867). We may rely on a theory of collections and collectives (60) and translate this example (Figure 5D) into:

- *rdf:type(everySwan, Collection)*
- *owl:SubClassOf(ncbitaxon:Cygnus, owl:SomeValuesFrom(member-of, owl:oneOf({everySwan})))*
- *owl:SubClassOf(owl:oneOf(everySwan), owl:AllValuesFrom(has-member, ncbitaxon:Cygnus))*

In other words, *'everySwan'* refers to a collection that has as its members all instances of the class Cygnus (NCBITaxon:8867), and that has only instances of that class as members.

We can use a similar translation to formalize other statements. An assertional statement unit (Fig. 6A) is translated directly into an ABox statement, and a universal statement unit follows a similar translation (Fig. 6D), resulting in the OWL ABox axiom:

*hasQuality(everySwan, someWhite)*.

Based on OWL semantics and the axioms to formalize *'everySwan'* and *'someWhite'* gives rise to the inference of the OWL axiom:

*owl: SubClassOf(ncbitaxon:Cygnus, owl:SomeValuesFrom(has-quality, White)*

as intended.

Similarly, contingent statement units (e.g., Figure 5B) are translated into standard ABox axioms, but prototypical statement units go a step further. They encode defaults while preserving the potential for exceptions—offering a formalism that integrates naturally with non-monotonic reasoning approaches. For instance, the prototypical statement *"Swans are typically white"* can be expressed as:

*hasQuality(x, white) :- hasQuality(swan, white), instanceOf(x, swan), not -hasQuality(x, white)*.

This captures the expectation that swans are white unless explicitly contradicted, thereby enabling flexible and defeasible reasoning within a unified semantic framework.

# Representing complex class axioms in FAIR knowledge graphs

**Challenge 3: Modelling class axioms with triangular relationships**
A persistent challenge in representing ontologies using RDF and OWL lies in the expression of **class axioms that involve triangular relationships**, i.e., relationships in which multiple parts of an axiom must refer to the same entity. This problem arises due to structural limitations in OWL and Description Logics, notably the use of **blank nodes** and the **tree model property** of Description Logics (61).

In OWL, universal statements about classes are typically defined in the TBox. When these are serialized in RDF, blank nodes are used to represent anonymous individuals (OWL:AnonymousIndividual), as shown in Figure 4C. While useful for defining composite structures, **blank nodes lack global identity**—they cannot be referenced outside or even within the same axiom in a meaningful way. This makes it difficult to express dependencies between different parts of an axiom that should share the same referent.

This limitation has severe implications for representing empirical knowledge, as blank nodes undermine the findability, accessibility, explorability, comparability, expandability, and reusability of any expression modelled as TBox, which is why empirical data should be preferably modelled as ABox expressions (for a detailed discussion see (62)).

A key example is the definition of a class *antenna type 1*, defined as "*An antenna that is longer than the eyes of the same organism that possesses it*". In Manchester Syntax (63), this axiom might be expressed as:

> 'has part some ((antenna and part of **some multicellular organism**) and has quality some (length and increased in magnitude relative to some (length and inheres in some (eye and part of **some multicellular organism**))))'.

Here, the phrase "*some multicellular organism*" appears twice. Under OWL semantics, each occurrence is interpreted as a distinct anonymous individual, due to the use of blank nodes. Consequently, the axiom is misinterpreted as comparing the antenna of one organism to the eye of the same *or* another, rather than *only* the same organism (64). This failure to co-refer results from the tree model constraint of Description Logics, which prohibits cyclic or triangular connections in class axioms.

**Approach: Semantic units for structured class axioms**
To address this limitation, we introduce a representational strategy based on three subcategories of compound units: item units, item group units, and class profile units. These allow us to model class axioms—including those involving triangular relationships—within the ABox, using the four novel representational resources rather than blank nodes, while retaining compatibility with OWL semantics through the abovementioned translations.

Item unit

An item unit is a compound unit comprising all statement units that share the same subject resource (27). Depending on the type of subject, item units can be assertional (named-individuals), contingent (some-instance), prototypical (most-instances), universal (every-instance), and all-instances (all-instances) item units. For example, all assertional statement units about a specific infected population—its location, basic reproduction number, and case fatality rate—form an assertional item unit.

Universal item units can be used to document simple **star-shaped class axioms** (see Fig. 7), defining **necessary properties** of every instance of a specific class. If the class axiom comprises **sufficient properties**, the respective semantic unit resource instantiates a **sufficient universal item unit**.

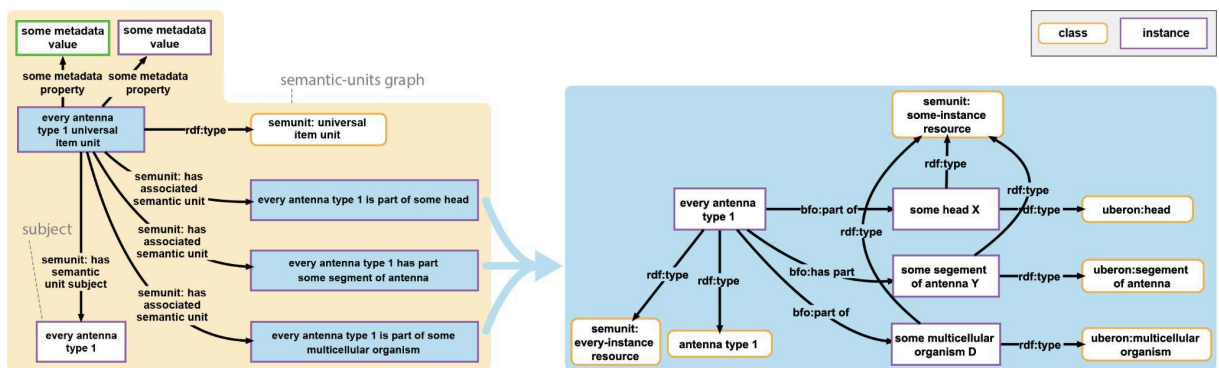

**Figure 7: Universal item unit. Left:** An example of a universal item unit, with three associated universal statement units. **Right:** The data graph of the same universal item unit, resulting from merging the data graphs of the three associated universal statement units. Each statement unit has *'every antenna type 1'* as their subject. *For reasons of clarity, neither resources and relations of associated semantic units nor their metadata shown.*

Item group unit

An item group unit is a compound unit formed by at least two item units whose data graphs are interconnected through intermediate statement units, where the object of one statement unit is the subject of another (27), allowing for **triangular** or **chain-like structures**.

**Universal item group units** are item group units that represent **class axioms** that cannot be expressed with a single universal item unit because they cover chains of relations comprising universal and contingent statement units with differing subject resources (Fig. 8). If a universal item group unit specifies sufficient properties for a class, it is a **sufficient universal item group unit**.

For instance, a universal item group unit can be used to represent a class axiom for the *antenna type 1* class from *Challenge 3*, representing the antenna and the eye as parts of the **same** organism (see Fig. 8, triangular relationship shown in red). By using some-instance and every-instance resources instead of blank nodes, the same (anonymous) organism can be referenced in multiple parts of the axiom. The resulting structure accurately captures the intended semantics of triangular relationships, and thus does not run into the restrictions of OWL.

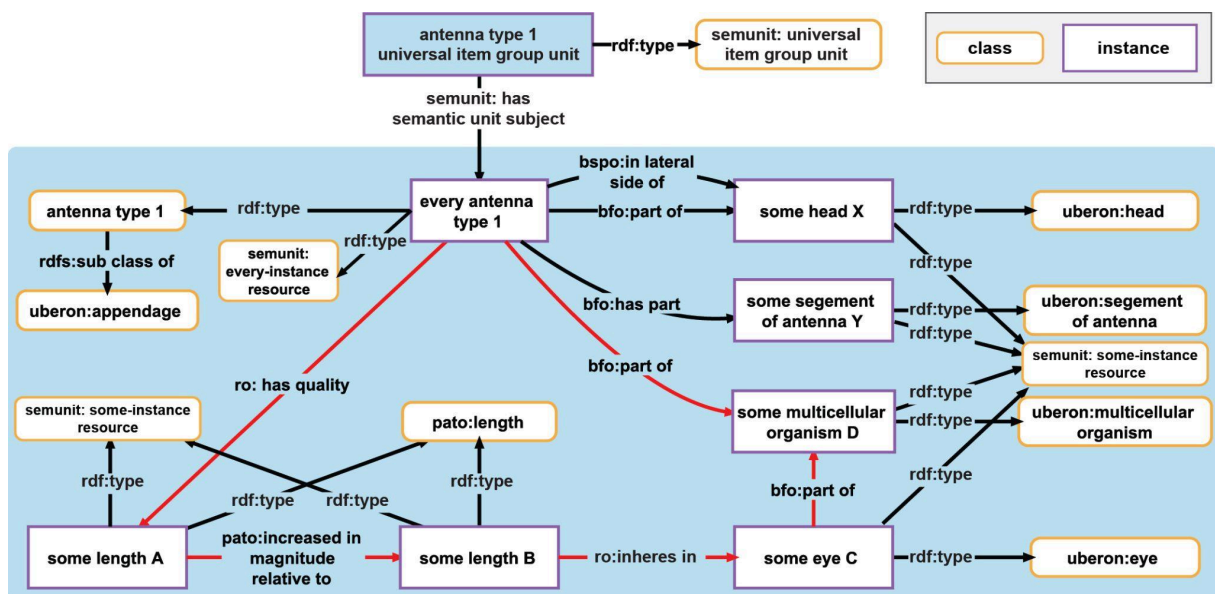

**Figure 8: Class axiom in a universal item group unit. A)** A universal item group unit that corresponds with the Manchester Syntax expression of the class axiom for *antenna type 1*, where multicellular organism (UBERON:0000468) is referenced twice, resulting in two independent blank nodes when turned into an OWL TBox expression (see *Challenge 3*). The data graph of this universal item group unit is shown in the blue box without borders (merged from the data graphs of all associated item and statement units). The statements cannot be modelled in a single universal item unit, because they do not all share the same subject. Contrary to the OWL TBox expression, this notation allows describing triangular relations as the one shown here in red, indicating that every instance of *antenna type 1* is longer than any eye that is part of the same organism. See also the correlation between the use of "some" in Manchester Syntax and the use of corresponding some-instance resources in the universal item group unit. *For reasons of clarity, neither resources and relations of associated semantic units nor their metadata are shown.*

The use of the logic programming paradigm allows us to formalize **complex statements** such as shown in Figure 8, which are **not directly expressible in OWL**, and the translation into OWL through translation patterns enables inferences within OWL when combined with other OWL axioms (53).

Class profile unit

While (sufficient) universal item units and (sufficient) universal item group units represent necessary and sufficient conditions for class membership, many real-world entity kinds are also characterized by **likely** and **contingent** properties. To capture this broader epistemic context, we define the **class profile unit**.

A class profile unit consists of the union of all universal, contingent, and prototypical item (group) units associated with a particular class. It serves as a **comprehensive representation of empirical and diagnostic knowledge**, going beyond formal class axioms to include statistical tendencies, typical features, conditional associations, and method-dependent recognition criteria for identifying members of the class (46).

For example, while a class axiom for *antenna type 1* may require it to be longer than the eye of the same organism (universal item group unit), the class profile may also include the fact that such antennas typically have a specific color, or frequently found in aquatic insects, even if these traits are not logically necessary.

# Representing negations and cardinality restrictions in FAIR knowledge graphs

**Challenge 4: Limitations of OWL in modelling negations and cardinality restrictions**
Scientific knowledge and data often involve not only affirmations of observed relationships but also **negations**, **absence statements**, and **cardinality restrictions**. While OWL-based ontologies can formally express these constructs, doing so often requires **complex class-level axioms**, making them difficult to represent, interpret, and query in practice—especially for non-experts. In contrast, labeled property graphs provide more intuitive mechanisms for expressing negations and cardinality constraints, but lack the formal semantics necessary for logical reasoning.

OWL, grounded in Description Logics, is built on the **Open World Assumption (OWA)**, which holds that the absence of a statement cannot be interpreted as the statement's negation. For example, if a knowledge graph does not state that a particular swan is white, we cannot infer that it is not white. Consequently, to express negations such as "*This swan is not white*", OWL requires the use of **class complements** in the TBox—rather than simple instance-level statements (ABox)—resulting in more complex and less cognitively interoperable representations (cf. 9A-D).

Similarly, **cardinality constraints**—e.g., stating that a swan has *exactly two wings*—must also be expressed through **class axioms**, which are again located in the TBox (Fig. 9E,F). While RDF/OWL offers the required expressivity to model these constraints, their use introduces a significant **cognitive burden** due to the need for logic-based formalisms and syntax.

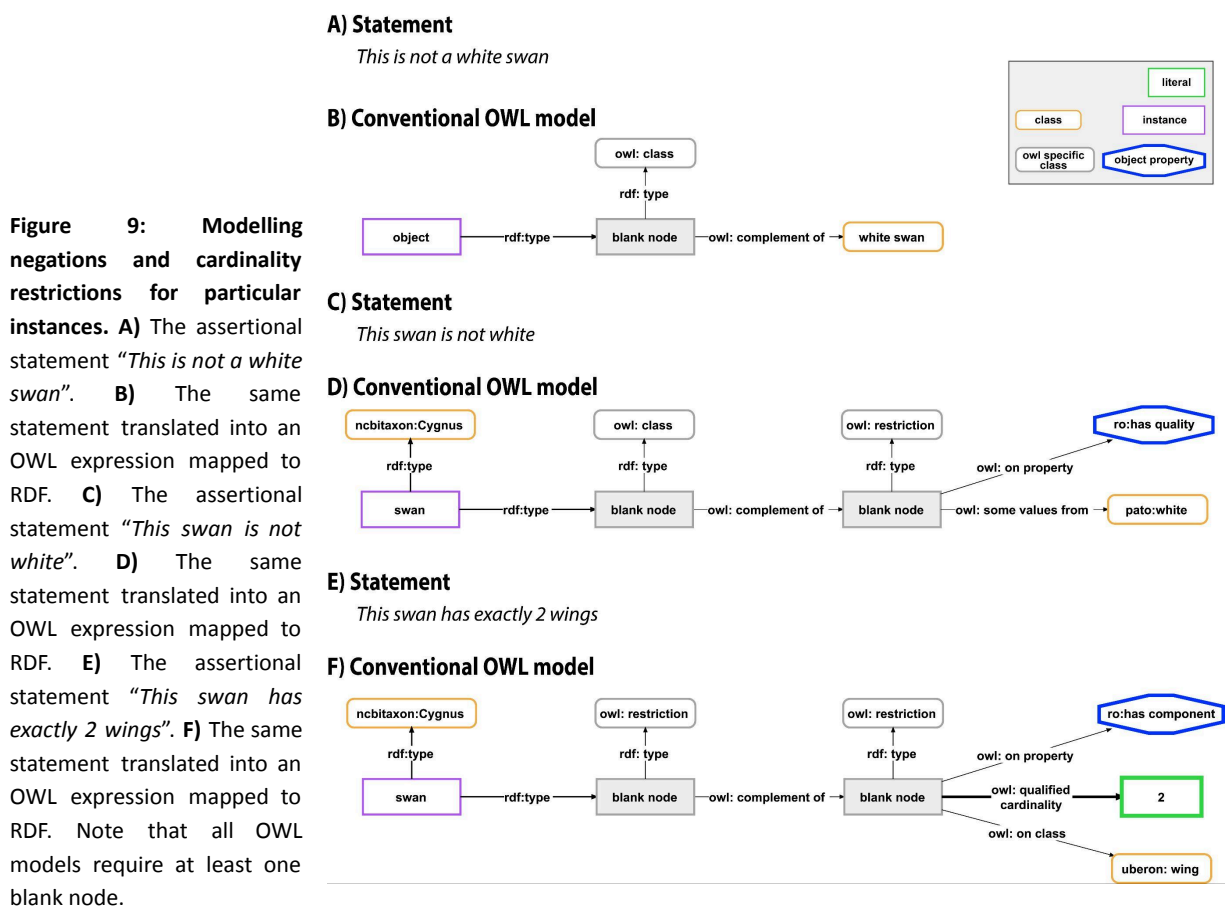

**Figure 9: Modelling negations and cardinality restrictions for particular instances. A)** The assertional statement "*This is not a white swan*". **B)** The same statement translated into an OWL expression mapped to RDF. **C)** The assertional statement "*This swan is not white*". **D)** The same statement translated into an OWL expression mapped to RDF. **E)** The assertional statement "*This swan has exactly 2 wings*". **F)** The same statement translated into an OWL expression mapped to RDF. Note that all OWL models require at least one blank node.

In sum, whereas labeled property graphs excel in usability and human-readability, thus offering a high cognitive interoperability, they lack formal semantics, while OWL excels in formal expressivity but suffers from complexity and poor accessibility for non-logicians. A modelling framework that bridges this gap must offer both human-intuitive and formally rigorous ways to represent negation and cardinality.

**Approach: Representing negation and cardinality with semantic units**

To address this challenge, we extend the semantic units framework by introducing two additional unit types: the **negation unit** and the **cardinality restriction unit**. These allow negation and quantitative constraints to be modelled directly in the ABox, making them easier to construct, query, and interpret—while still permitting translation to OWL TBox axioms when needed.

Representing negation in the ABox

OWL expresses **negation** through class complements. For example, the statement "*This fruit is not a pome fruit*" corresponds to a class expression of the form '*not (type pome fruit)*' in Manchester Syntax (Fig. 10A,B). This requires defining the individual as a member of the class fruit (PO:0009001) and also as a member of the class that is the complement of pome fruit (PO:0030110) (Fig. 10C).

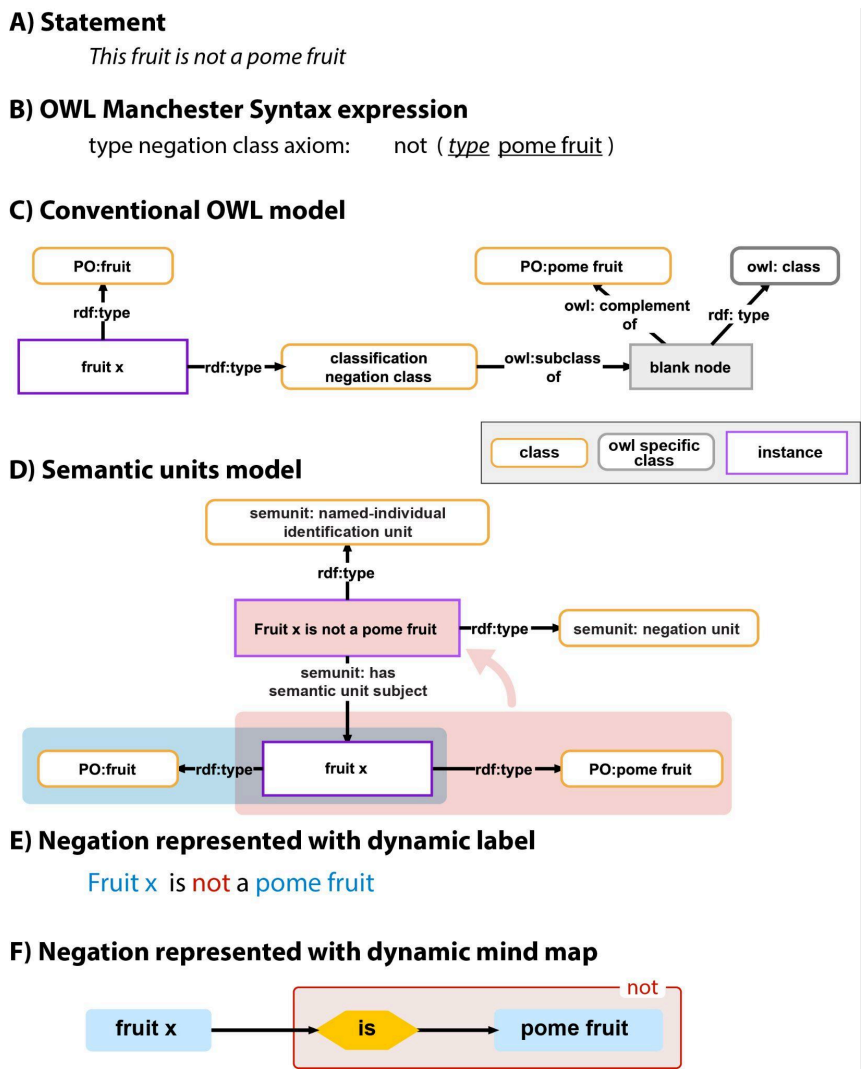

**Figure 10: Modelling negations involving instances by applying semantic units. A)** A human-readable statement that this fruit is not a pome fruit. The statement can be modelled in two different ways. As an OWL expression that can be specified using **B)** Manchester Syntax, where the fruit is an instance of a class that is defined as all of its instances are not instances of pome fruit (PO:0030110). Note, how this Manchester Syntax expression translates into **C)** an OWL expression mapped to RDF, where 'fruit x' is an instance of fruit (PO:0009001) but also of a class that is the complement to pome fruit (PO:0030110). Alternatively, the statement can be modelled as an instance-based graph using semantic units **D)**. The data graph in the blue box states that the entity is a fruit. The data graph belonging to the statement unit stating the negation (red box with borders, here shown with its dynamic label), on the other hand, is shown in the red box without borders and states that the entity is a pome fruit. However, since the latter statement unit not only instantiates the class *SEMUNIT: named-individual identification unit* but also *SEMUNIT: negation unit*, it actually negates the statement in the data graph, therewith indicating that it is *not* a pome fruit. Utilizing the UI display patterns of the statement unit, the graph can be displayed in the UI of a knowledge graph in a human-readable form, either as text through a dynamic label **E)** or as a graph through a dynamic mind map **F)**. *For reasons of clarity, metadata for each semantic unit is not represented.*

Using semantic units, this same negation can be modelled more intuitively. The statement is represented by two units:

1. A **named-individual identification unit** for the fruit in question.
2. Another **named-individual identification unit** whose subject is the fruit and whose object is the class pome fruit (PO:0030110), but which is also an instance of the *SEMUNIT: negation unit* (see Fig. 10D).

By tagging the statement unit as a negation unit, the statement is inverted in meaning, eliminating the need for TBox-level constructs. For UIs, display patterns can render such statements in human-readable forms (Fig. 10E,F).

### Representing absence in the ABox

Negation also arises in absence statements. The claim "*This head has no antenna*" is logically equivalent to the OWL axiom '*type: not (has part some antenna)*', which, again, requires class complement expressions (Fig. 11A-C).

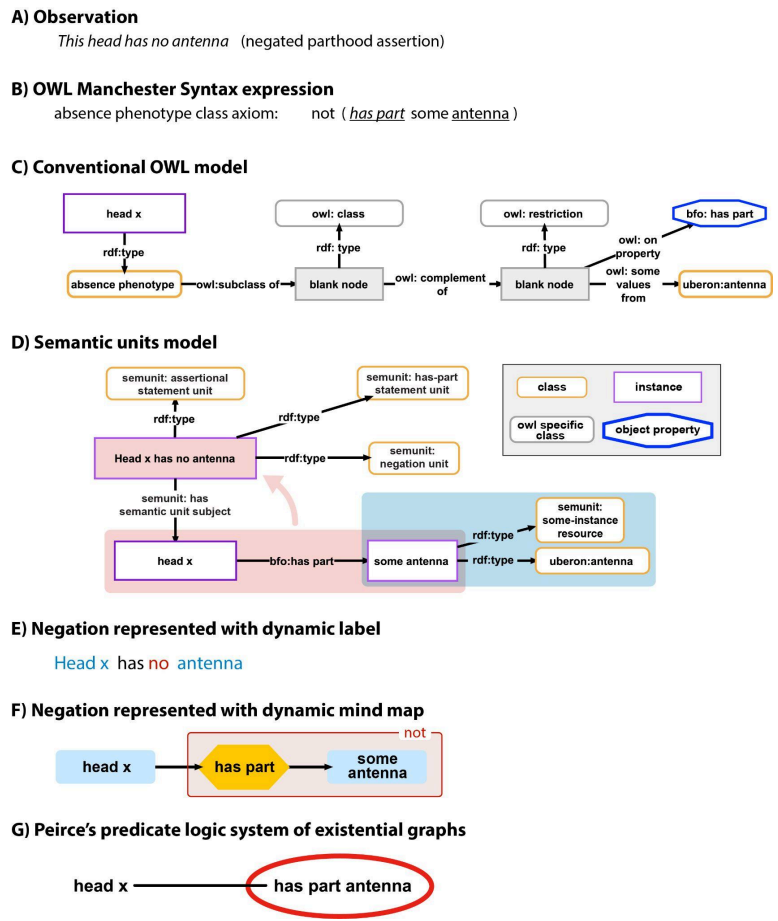

**Figure 11: Relation between an absence observation, the corresponding assertions, and two alternative ways to model them in a knowledge graph**. **A)** A human-readable statement about the observation that a given head has no antenna. **B)** In OWL, absence statements cannot be expressed as relations between instances. Therefore, the observation from A) must be expressed using a class expression, which can be formulated using Manchester syntax. Following this notation, the head would be an instance of a class that is defined to have only instances that have no antenna as their parts ('not' and 'some' being used as mathematical expressions). **C)** The translation of the assertion from A) and B) into an OWL expression mapped to RDF. Note how *absence phenotype* is defined as a set of relations of subclass and complement restrictions involving two blank nodes. **D)** The same statement can be modelled using two semantic units. One of them is modelling the has-part relation and negates it (red box with borders and its dynamic label, with its data graph in the red box without borders). It is therefore an instance of *SEMUNIT: has-part statement unit* as well as *SEMUNIT: assertional statement unit* and *SEMUNIT: negation unit*.

The other semantic unit is an instance of *SEMUNIT: some-instance identification unit* and relates 'some antenna' to antenna (UBERON:0000972) and *SEMUNIT: some-instance resource* via the property *type* (RDF:type). Its data graph is shown in the blue box. Together, they model the observation from A). Utilizing UI display patterns of the statement unit, the graph can be displayed in the UI of a knowledge graph in a human-readable form, either as text through a dynamic label **E)** or graphically through a dynamic mind map **F)**. **G)** An alternative notation of the statement 'this head has no antenna' and the observation from A). The notation uses Peirce's predicate logic system of existential graphs. The *identity line* ── between the two phrases 'head x' and 'has part antenna' states that head x has some antenna as its part, whereas the red circle surrounding the latter phrase expresses its negation by crossing the *line of identity*. *For reason of clarity, the relation between 'head x' and head (UBERON:0000033) is not shown in C) and D). Moreover, metadata for the semantic unit is not represented.*

In the semantic unit framework, this is expressed as follows (Fig. 11D):

- A *SEMUNIT: has-part statement unit* is created, linking the 'head x' (UBERON:0000033) to a some-instance resource of type antenna, i.e., 'some antenna' (UBERON:0000972).
- This statement unit is also typed as *SEMUNIT: negation unit*.
- The some-instance resource is modelled using a *SEMUNIT: some-instance identification unit*.

In general, whenever a relation involving a type of entity (e.g., antenna (UBERON:0000972)) is asserted as **absent**, the statement unit is simply typed as an instance of *SEMUNIT: negation unit*. As before, UI display patterns can be used to render this information in a user-friendly manner (Fig. 11 E,F).

## Representing negation of relations between individuals

The framework also supports negation of relationships between two specific instances (e.g., "*This fruit is not part of this orange plant*"), corresponding to OWL's negative property assertion (OWL:negativePropertyAssertion) (Fig. 12). The negated statement is again expressed by classifying the relevant statement unit as an instance of *SEMUNIT: negation unit*.

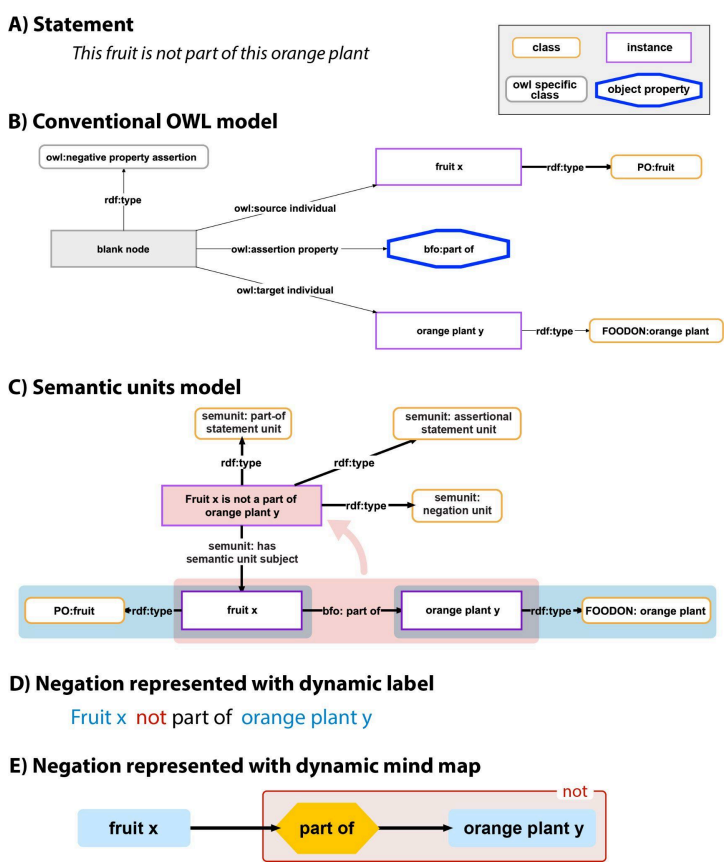

**Figure 12: Modelling negating relations between instances by applying semantic units.** **A)** A human-readable statement that this fruit is not part of this organge plant. The statement can be modelled in two different ways: **B)** as an OWL expression mapped to RDF. Note, how the statement is translated into a negated assertion statement with source, property, and target specification, relating an instance 'fruit x' (PO:0009001) to an instance 'orange plant y' (FOODON:03411339) involving a blank node; or **C)** using three semantic units. The two named-individual identification units, with their data graphs shown in the blue boxes, state that one of the objects is a fruit and the other one an orange plant, whereas the part-of statement unit (red bordered box and its dynamic label, with its data graph shown in the red box without borders) states that fruit x is not part of orange plant y as the unit instantiates both *SEMUNIT: part-of statement unit*, *SEMUNIT: assertional statement unit*, and *SEMUNIT: negation unit*.

Utilizing the UI display patterns of the statement unit, the graph can be displayed in the UI of a knowledge graph in a human-readable form, either as text through a dynamic label **D)** or graphically as a dynamic mind map **E)**. *For reasons of clarity, metadata for the semantic unit is not represented.*

### Peirce's existential relational graphs

The modelling approach for negation implemented in the semantic units framework is philosophically grounded in Peirce's **existential relational graphs** (65,66), where a *line of identity* '—' denotes that '*something is A*' (—$A$) (Fig. 11G). The *line of identity* can be understood to represent an existential quantifier ($\exists x$). The interruption of this line with a circle enclosing $A$ denotes negation, i.e., '*something is not A*' ($\neg A$). The resulting existential relational graphs are sufficiently general to represent full First-Order Logic with equality (66), and conceptually mirror our use of the negation unit.

### Representing cardinality restrictions

Expressing **cardinality restrictions** in OWL requires the use of class expressions that indicate the cardinality as a class restriction. Instead of using this rather complex class axiom to describe for instance that a given head possesses exactly three eyes (Fig. 13B), following the semantic unit framework, the same information could be modelled by (Fig. 13C):

- Creating a has-part statement unit linking the 'head x' to a some-instance resource 'some eye' (UBERON:0000970).
- 'some eye' (UBERON:0000970) is the object of this relation and is defined in a *SEMUNIT: some-instance identification unit* that is also a *SEMUNIT: cardinality restriction unit*.
- A numeric value (e.g., 3) is linked to 'some eye' within the *SEMUNIT: some-instance identification unit* via the property *qualified cardinality* (OWL:qualifiedCardinality).

This pattern allows cardinality to be expressed as an instance-level quantitative constraint, making it easier to query and visualize. Furthermore, ranges and frequencies can be modelled using float values combined with appropriate units (e.g., count unit (UO:0000189) or percent (UO:0000187)).

UI display patterns can render this information as readable statements (Fig. 13D,E).

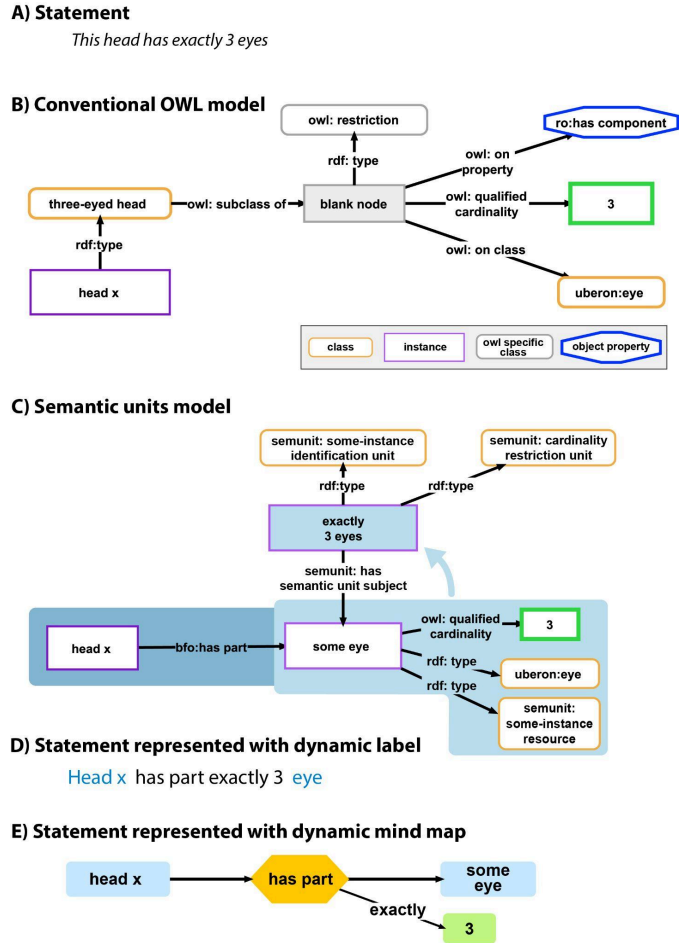

**Figure 13: Modelling cardinality restrictions involving instances by applying semantic units**. **A)** A human-readable statement that this head has exactly three eyes. The statement can be modelled in two different ways: **B)** as an OWL expression mapped to RDF. Note, how 'has exactly three eyes' is translated into being an instance of a class that has a cardinality restriction on the <u>has component</u> property (RO:0002180) and the class <u>eye</u> (UBERON:0000970) with a cardinality value of 3, thereby involving one blank node; or **C)** using two semantic units, one of which models the has-part relationship, with its data graph shown in the dark blue box. The other semantic unit (light blue bordered box and its dynamic label, with its data graph shown in the light blue box without borders) instantiates *<u>SEMUNIT: assertional statement unit</u>*, *<u>SEMUNIT: some-instance identification unit</u>*, and *<u>SEMUNIT: cardinality restriction unit</u>*. Utilizing the UI display patterns of the statement unit, the graph can be displayed in the UI of a knowledge graph in a human-readable form, either as text through a dynamic label **D)** or graphically as a dynamic mind map **E)**. *For reasons of clarity, metadata for the semantic unit is not represented.*

**Formal semantics and OWL translation for negation units and cardinality restriction units**
Although reasoners that natively support semantic units do not yet exist, the structures we define can be translated into OWL axioms using **logic programs**, as proposed in (53).

For example, a cardinality restriction unit can be translated as follows:

*cardinality-restriction-unit(x), has-semantic-unit-subject(x,y), owl:qualified-cardinality(y,z), some-instance-of(y, w).*

Applying this general logic program pattern to the cardinality restriction example from Figure 13C, it corresponds to the OWL axiom:

*rdf:type(cX, owl:intersectionOf(Collection, owl:cardinality(has-member, 3, uberon:eye)))* ,

where *cX* is a new individual name (i.e., a name not used elsewhere). This will then be combined with the '<u>part-of assertional statement unit</u>' for '<u>head X</u>' in the same example to assert the ABox statement:

*part-of(`head X', cX).*

Similarly, we can model (classical) **negation** by specifying negated assertional statement units (Figure 10) via the rule:

*NegatedStatementUnit(x) :- NegationUnit(x), StatementUnit(x)*.

Using the following logic program as precondition:

*NegatedStatementUnit(x), has-semantic-unit-subject(x,y), rdf:type(y,z)*,

this translates to OWL as:

*rdf:type(y, owl:complementOf(z))*.

This example shows the flexibility of our use of logic programming, as the *NegatedStatementUnit* predicate is inferred from the two assertions added to the statement unit and does not have to be asserted. However, because multiple patterns may apply to statements, to avoid inconsistencies, translation of a standard statement should be suppressed if a negation unit is present, using a weak negation in the logic program:

*StatementUnit(x) :- not NegationUnit(x)*.

This approach shows how negation and cardinality restrictions, while difficult to model in OWL due to reliance on class-level expressions, can be represented more naturally and intuitively using semantic units. By introducing negation units and cardinality restriction units, we enable instance-level modelling that remains semantically faithful and translatable to OWL.

From a scholarly perspective, this method enhances both cognitive interoperability and machine-actionable semantics, offering an intermediate solution that balances usability with formal rigor. It simplifies querying and visual representation, aligns with philosophical models of negation (e.g., Peirce's existential relational graphs), and opens pathways for new types of reasoning in hybrid graph environments.

# Representing epistemic beliefs and disagreement in FAIR knowledge graphs

### Challenge 5: Modelling epistemic beliefs and disagreement

Scientific knowledge production inherently involves **epistemic stances**—expressions of certainty, doubt, disagreement, and agnosticism regarding assertions. For example, one researcher (*Person A*) may claim "*All swans are white*", another (*Person B*) may explicitly reject this, and a third (*Person C*) may remain agnostic. Further complexity arises when scientists express beliefs about other's beliefs (e.g., *Person A asserts that Person B holds the belief that all swans are white*) or explicitly agree or disagree with other's epistemic positions.

Capturing such nuanced forms of **epistemic attribution**, including **disagreement**, **uncertainty**, and **belief about beliefs**, is essential for accurately representing the evolving structure and dynamics of scientific knowledge and scholarly discourse. However, **standard semantic web technologies—both OWL ontologies or knowledge graphs and labeled property graphs—lack native constructs or formal semantics for such epistemic modelling**. OWL primarily supports factual statements under OWA but does not incorporate modalities of belief, doubt, or disagreement. Similarly, property graphs excel in flexible relational modelling but do not offer formal means to represent or reason over epistemic states.

As a result, documenting epistemic beliefs or their attribution within knowledge graphs remains a significant challenge, often requiring bespoke representational extensions or external frameworks beyond existing standards.

**Approach: Modelling epistemic beliefs and disagreement using semantic units**

To address this gap, we propose an extension of the semantic unit framework by introducing **epistemic units**—a specialized class of complex statement units that explicitly associate a person or agent with a statement, thereby encoding that person's **epistemic stance** regarding that statement.

Our framework defines a taxonomy of epistemic units, comprising three primary categories, each representing a fundamental epistemic attitude:

1. **Positive epistemic unit**: Models affirmation or belief in a statement (e.g., *Person A* believes "*This fruit is a pome fruit*") (Fig. 14A).
2. **Negative epistemic unit**: Represents disbelief or rejection of a statement (e.g., *Person B* rejects "*This fruit is a pome fruit*") (Fig. 14B).
3. **Agnostic epistemic unit**: Encodes agnosticism or uncertainty regarding a statement (e.g., *Person C* neither affirms nor denies "*This fruit is a pome fruit*") (Fig. 14C).

Each epistemic unit links the agent (person) to the relevant statement unit, which itself has a GUPRI. This structure enables straightforward querying: retrieving all positive, negative, or agnostic epistemic units referencing a given statement provides a clear map of who agrees, disagrees, or is uncertain about that statement.

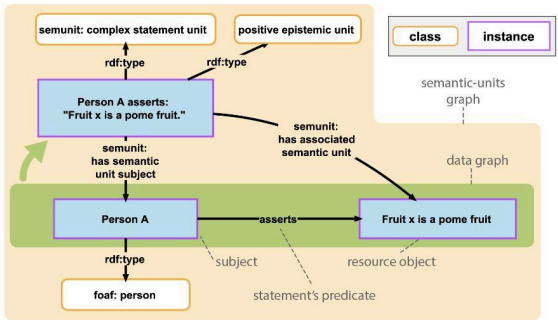
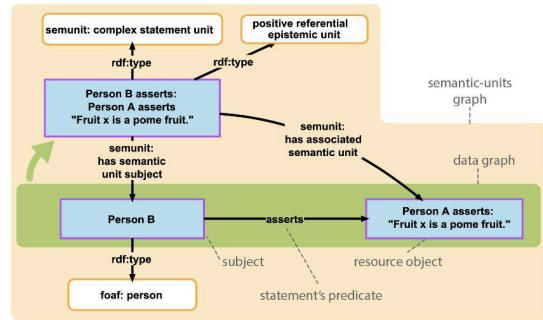
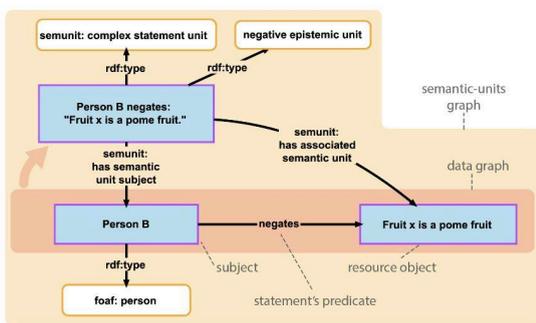
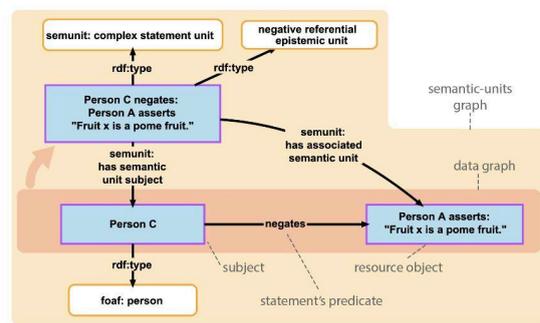
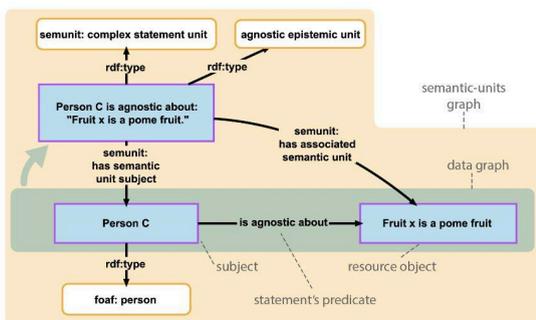
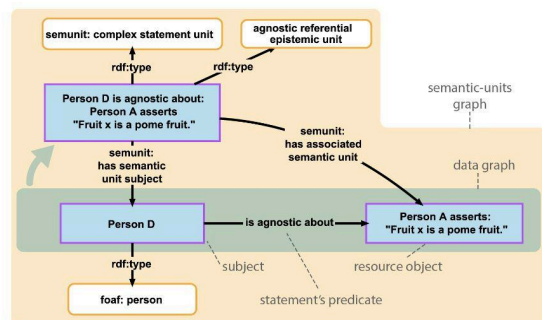

**Figure 14: Modelling epistemic beliefs and disagreement by applying semantic units.** Six examples of different types of epistemic units. *For reasons of clarity, metadata for each semantic unit is not represented.*

Referential epistemic units: Modelling beliefs about beliefs

Beyond direct epistemic stances, scientific discourse frequently involves **epistemic meta-attributions**, where an agent holds a belief regarding another agent's belief. For example, *Person B* may hold the belief that *Person A* believes "*This fruit is a pome fruit*". Such **referential epistemic beliefs** require a more expressive representational construct.

We model these using **referential epistemic units**, which are analogous to the primary epistemic units but differ in that their **object positions are themselves epistemic units** (Fig. 14D-F). Referential epistemic units can be:

1. **Positive referential epistemic units**: *Person B* asserts that *Person A* believes "*This fruit is a pome fruit*" (Fig. 14D).
2. **Negative referential epistemic units**: *Person C* denies that *Person A* believes "*This fruit is a pome fruit*" (Fig. 14E).
3. **Agnostic referential epistemic units**: *Person D* is uncertain about *Person A*'s belief in "*This fruit is a pome fruit*" (Fig. 14F).

This recursive structure enables rich modelling of **second-order epistemic attitudes** and facilitates capturing the full complexity of scientific debate, consensus formation, and disagreement.

**Extending epistemic modelling through logic programming**

In addition to the representational benefits of semantic units, the l**ogic programming paradigm** enables a formal bridge between statement units (treated as individuals) and their propositional content (53). Each epistemic unit—such as those shown in Figure 14—is not only a standalone instance (an *epistemic statement unit*) but also serves as a referential anchor in a logic program. These statement units can be translated into OWL axioms via the transformation patterns described earlier. Moreover, by treating epistemic statement units as individuals within logic programs, **relations between statements themselves—such as belief, contradiction, or support—can be explicitly modelled**.

This dual-level modelling enables the use of **answer set programming** for expressing and reasoning over **argument structures**—for instance, to detect conflicts between beliefs, resolve inconsistencies, or generate alternative belief sets under different assumptions (67,68). In this way, semantic units do not merely serve as a static knowledge representation mechanism but can also potentially participate in **dynamic**, **rule-based reasoning environments**, supporting applications such as scientific debate modelling, provenance analysis, and epistemic network visualization.

# Findability, discoverability, and querying data and knowledge in FAIR knowledge graphs

**Challenge 6: The complexity of querying knowledge graphs**

Interacting with a knowledge graph—whether to add, retrieve, update, or delete data—typically necessitates proficiency in **graph query languages** such as [SPARQL](#) for RDF/OWL-based graphs or [Cypher](#) for labeled property graphs like Neo4j. While these languages are powerful, they pose significant barriers to entry. Crafting effective queries demands not only familiarity with the specific syntax but also a deep understanding of the schemata underlying the graph to be queried, a combination of skills that many users, including developers, may lack. Even technically proficient individuals often find the process time-consuming and error-prone. This complexity hinders the broader adaption and seamless utilization of knowledge graphs (69).

Recent advancements in **LLMs** have shown promise in **translating natural language questions into SPARQL queries** (9). However, studies indicate that while LLMs can handle simple queries with reasonable accuracy, their performance degrades with increasing query complexity. Natural language inputs are often ambiguous, leading to incorrect or unintended query results when interpreted by an LLM (10). Furthermore, current LLMs struggle to reliably handle complex or deeply nested query structures, reducing their effectiveness in more advanced use cases (11). Whether future advancements in LLM architectures and training methodologies will overcome these limitations remains an open question, underscoring the need for alternative approaches that can bridge the gap between user intent and formal query construction.

**Challenge 7: Representing interrogative statements in knowledge graphs**
**Interrogative statements**, or **questions**, are fundamental to information retrieval. In the context of knowledge graphs, transforming these natural language questions into formal queries is essential for machine-actionable operations. However, current knowledge graph formalisms, including OWL and labeled property graph models like Neo4j, lack native support for representing interrogative statements within the knowledge graph. Moreover, there is no standard mechanism to formally distinguish between different types of questions within a graph-based framework. Consequently, interrogative statements remain outside the formal domain of discourse in knowledge graphs, analogous to the challenges posed by representing universal statements in knowledge graphs discussed earlier (*Challenge 2*).

**Approach: Modelling interrogation statements as question units**
To address these challenges, we introduce the concept of **question units**—a novel category of semantic units designed to represent interrogative statements within knowledge graphs—and combine them with the new representational resource types we introduced earlier. This approach leverages existing semantic unit classes as sources, rephrasing them into interrogative forms that can be documented within the graph. By doing so, searches become objects within the knowledge graph (comparable to (70,71)), facilitating more intuitive and accessible querying mechanisms.

For instance, considering the assertional statement "*Apple X has a weight of 204.56 grams*" (Figure 1A). This source statement can be transformed into the question "*Does apple X have a weight of 204.56 grams?*" by copying the data graph of the source semantic unit (Fig. 1B) and classifying it as a question unit (Fig. 15A). When translated into a query, a Boolean *true/false* would be returned.

Further modifications, such as **underspecifying** a subject or object resource or literal, allow for the formulation of more generalized questions. The weight value object literal from the above example statement can for instance be underspecified by defining a value range (e.g., '*xsd:float[>=0.0f]*'), resulting in the questions "*What is the weight of apple X?*" (Fig. 15B). The question "*Which apple has a weight of 200 to 300 grams?*" (Fig. 15C) not only requires setting the weight value to a value range (i.e., or '*xsd:float[>=200.0f, <=300.0f]*') but also underspecifying the subject resource by replacing the named individual resource '<u>apple X</u>' with a some-instance resource of <u>apple</u> (NCIT:C71985). These transformations enable the representation of various interrogative forms within a knowledge graph.

In correspondence with the type of statement unit that serves as the source of the question, we can distinguish basic types of question units (Fig. 16).

The creation of a semantic unit, accompanied by the underspecification of one or more slots in its data graph, facilitates the formation of **question units that can be stored in a knowledge graph** (see *Challenge 7*). Literal slots are underspecified by for instance specifying a value range, while resource slots are underspecified by replacing a named-individual resource with a some-instance resource of a specific class, thereby specifying that the answer to the question may have any instance of that class in that slot. This approach is versatile, as it can be applied to any type of statement unit class as its source.

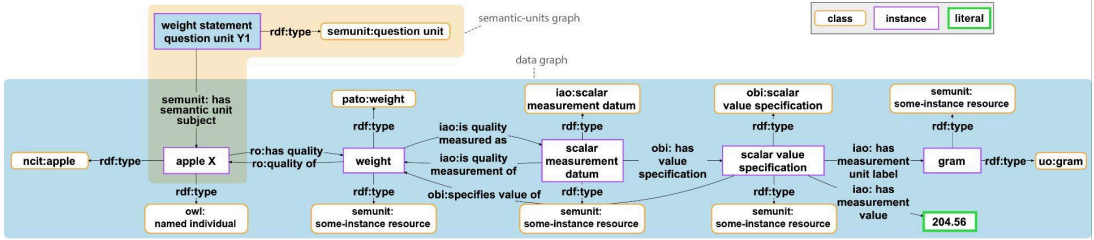
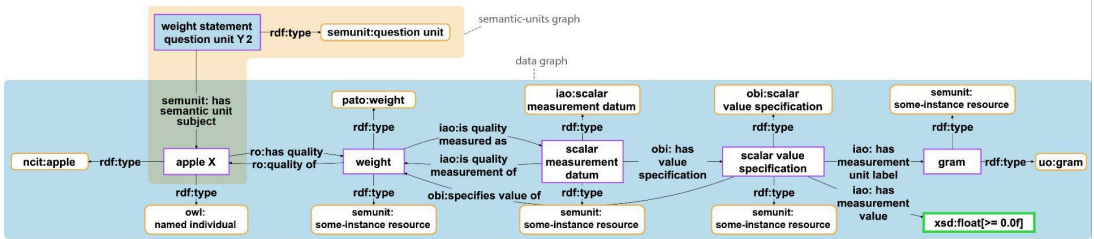
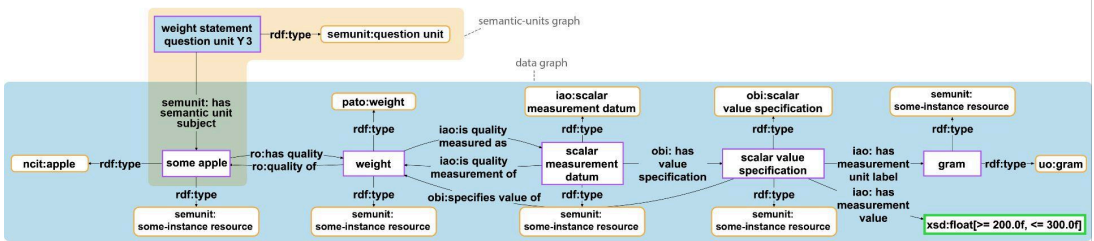
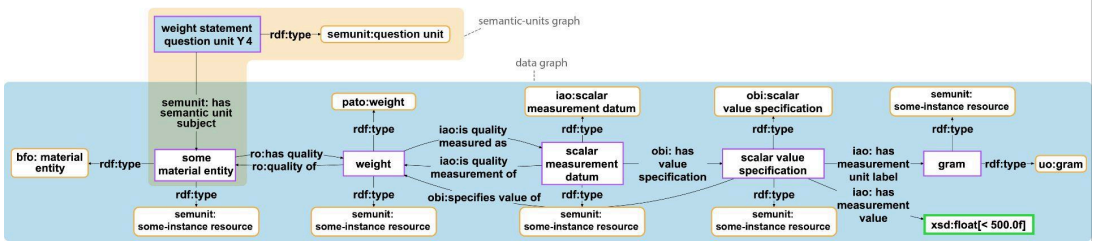

**Figure 15: Question units**. Four examples of question units. *For reasons of clarity, metadata for each semantic unit is not represented.*

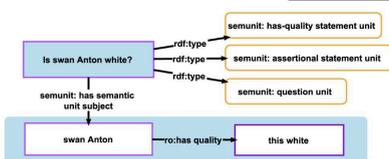
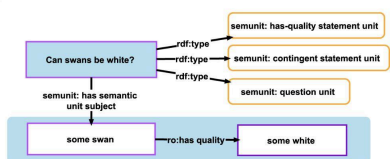
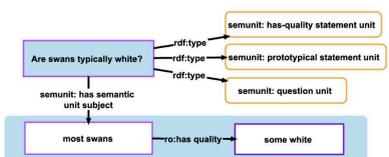
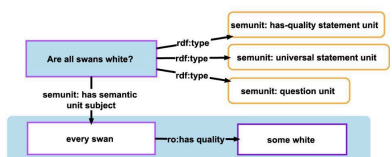

**Figure 16: Modelling interrogation statements as question units**. Modelling the four interrogation statements from *Challenge 7* as question units. *For reasons of clarity, metadata for each semantic unit is not represented.*

The integration of question units with SHACL shapes facilitates the development of **query-builders** that can translate question units into executable queries. By utilizing SHACL shapes, we can derive input forms corresponding to the structure of the source statement unit, allowing users to construct queries without requiring familiarity with complex query languages (see *Challenge 6*). The use of Boolean operators with **Boolean units** and the reusability of object resources as subject resources in another question unit facilitates the combination of multiple statement-based question units to form a more complex **compound question unit**. Depending on the Boolean operator, the following Boolean units can be distinguished: **Boolean AND**, **Boolean OR**, **Boolean XOR**, **Boolean NOT**, and **Boolean EQUAL units**. It is imperative to note that all resources associated with a Boolean unit are interconnected through the unit's designated Boolean operator.

This approach not only enhances accessibility but also supports the documentation of competency questions within the graph, providing a mechanism for ongoing evaluation and refinement of the knowledge graph.

## Standard views on data and knowledge in FAIR knowledge graphs

**Challenge 8: Extracting topic-specific subgraphs for standardized views**
In many domains, the description and exchange of information are governed by standardized formats that specify mandatory and optional content elements. For example, in the manufacturing and automotive industries, material data sheets—such as those following the [International Material Data System](#) (IMDS)—specify required information about a material's physical properties, compositional elements, and regulatory status. These data sheets are passed between suppliers and customers, embedded into compliance workflows, and used as authoritative documentation.

Within organizational knowledge graphs, it becomes increasingly important to replicate such structured, standardized views of information. In the context of knowledge graphs, this requires the ability to extract and represent domain-specific subgraphs—each corresponding to a well-defined information schema—within a broader graph structure. These subgraphs function as modular, referable content blocks that encapsulate the relevant statements describing a specific topic or entity in a standardized way.

More broadly, the capacity to define, reference, and retrieve such **topic-specific subgraphs**—each representing a logically coherent, contextually meaningful view over the graph—can greatly enhance modularity, data reuse, and interoperability. If each subgraph is given its own GUPRI, it becomes a referable unit of information. This enables not only machine-readable reuse across contexts but also the ability to make further statements *about* these information bundles (e.g., provenance, trust, or evaluation metadata). This form of modularization addresses a key usability gap in RDF/OWL and labeled property graph systems, which lack built-in mechanisms for **grouping and referencing semantically coherent collections of statements as first-class entities** and for **contextual graph exploration** (25).

**Approach: Representing standardized views as standard information units**
To address this challenge, we introduce the concept of **standard information units**—a type of compound unit that groups together statement units into a coherent, context-specific subgraph representing a standardized view over some domain of knowledge. Each standard information unit encapsulates a set of semantically related statements and provides a dedicated resource (GUPRI) for referencing the collection as a whole.

Standard information units enable explicit modelling of canonical data structures within the graph. Examples include:

- A standard material data sheet in engineering, capturing mandatory and optional statements about a substance's physical, chemical, and regulatory attributes.
- A standardized product specification sheet in retail or e-commerce contexts.
- The collection of all statement units in the graph that model the scientific findings of a particular scholarly publication.
- A user profile aggregating public-facing personal data within a platform or application.

Which specific types of statement units constitute a given standard information unit are determined by the context and shaped by the expectations or requirements of the relevant domain.

Moreover, standard information units often characterize or are associated with a specific entity in the graph (e.g., a particular material, product, or user). This relationship is formally represented within the standard informatio unit through the property *<u>SEMUNIT: has associated standard information unit</u>*, linking the subject entity to the unit that semantically describes it.

By formalizing these structures, standard information units make it possible to:

- Identify and retrieve semantically coherent subgraphs on demand.
- Create views aligned with domain-specific information expectations and their corresponding schemata.
- Reuse and cite information collections with GUPRI-based referencing.
- Enable modular graph construction, content packaging, and workflow integration.

These capacities position standard information units as a foundational construct for structuring and modularizing knowledge graphs in a scalable and interoperable manner, aligning with **FAIR Principles** by promoting **findability** and **reusability** through named and structured units of knowledge.

# Representing geo-indexed and temporally ordered data in FAIR knowledge graphs

**Challenge 9: Modelling spatio-temporal and sequential information**
In many scientific, historical, and real-world domains, the truth value and relevance of statements are intrinsically bound to *temporal* and *geographical* contexts. For instance, the assertion that John F. Kennedy was the 35$^{th}$ President of the United States holds only within a specific historical time (1961-1963) and implies a particular position within a temporal sequence (the U.S. presidential timeline). Similarly, a statement such as "*The G7 Summit took place in Fasano, Italy*" is meaningful only when anchored to a particular date, location, and iteration in the series of summits.

Such spatio-temporal indexing becomes especially critical in domains dealing with events, processes, and episodic phenomena, including:

- Scientific workflows and experimental protocols (e.g., multi-phase lab procedures),

- Clinical interventions (e.g., patient treatment sequences),
- Historical reports,
- Environmental phenomena (e.g., seasonal weather anomalies tied to geospatial zones).

Effectively representing these time- and location-specific claims within RDF/OWL-based knowledge graphs, however, remains technically challenging, especially when a statement is modelled across multiple interconnected triples. OWL itself lacks first-class support for temporal or spatial constructs, offering no native operators for modelling time intervals, geo-coordinates, or ordered sequences. RDF allows such representations via reification, named graphs, or singleton properties, but these approaches introduce overhead, reduce clarity, and often lead to interoperability issues due to divergent implementation practices. Moreover, RDF's triple-based structure complicates the modelling of metadata about statement *groups*, which is essential when spatio-temporal qualifiers span across multiple linked assertions.

**Approach: Modelling time-indexed, geo-indexed, and time-ordered units**
To address this challenge, we propose three specialized assertional statement unit types—**time-index**, **geo-index**, and **time-order statement units**—and three specialized compound unit types—**time-indexed units**, **geo-indexed units**, and **time-ordered units**—that extend the semantic unit model to explicitly capture and contextualize temporal, spatial, and sequential dimensions of knowledge.

- **Statement unit subtypes:**
    - **Time-index statement units** specify a specific temporal context. This can be a specific point in time (e.g., a timestamp) or a time interval (e.g., a presidential term, a conference duration).
    - **Geo-index statement units** specify geospatial metadata such as GPS coordinates, administrative regions, or named locations.
    - **Time-order statement units** specify a position within an ordered timeline or procedural narrative.

- **Compound unit subtypes:**
    - **Time-indexed units** encapsulate statement units that are valid only within a specific temporal context. They always include some time-index statement unit.
    - **Geo-indexed units** contain information that is dependent on some location, and thus always include some geo-index statement unit.
    - **Time-ordered units** cover statements that refer to a specific position within a sequential order. They always include some time-order statement unit that allows the construction of ordered timelines or procedural narratives.

These units allow spatio-temporal qualifiers to be embedded directly into the semantic unit structure of the knowledge graph, while preserving modularity and referential integrity.

For instance, consider modelling the presidency of John F. Kennedy (Fig. 17A):

- An assertional statement unit specifies that John F. Kennedy was president of the U.S.,
- A time-index statement unit specifies the duration of the presidency (1961-1963),
- A time-order statement unit his position in a historical sequence of U.S. presidents,

- A complex statement unit aggregates these together to form a cohesive narrative about his presidency, instantiating both a time-indexed unit and time-ordered unit.

Similarly, modelling the 50th G7 Summit includes (Fig. 17B):

- A geo-index statement unit for the venue in Fasano, Italy,
- A time-index statement unit for the dates of the summit,
- A time-order statement unit for its position in a historical sequence of G7 Summits,
- An item group unit linking this event to the broader G7 series,
- A complex statement unit aggregates these together to form a unit of information, instantiating simultaneously a geo-indexed unit, a time-indexed unit, and a time-ordered unit.

This structured approach enables precise querying, reasoning, and presentation of information with respect to time and place. It also ensures compatibility with timeline visualizations, location-based filtering, and process modelling applications.

By embedding temporal and spatial context into the graph at the unit level—rather than resorting to ad hoc metadata overlay—the approach supports semantic clarity, machine-actionability, and interoperability.

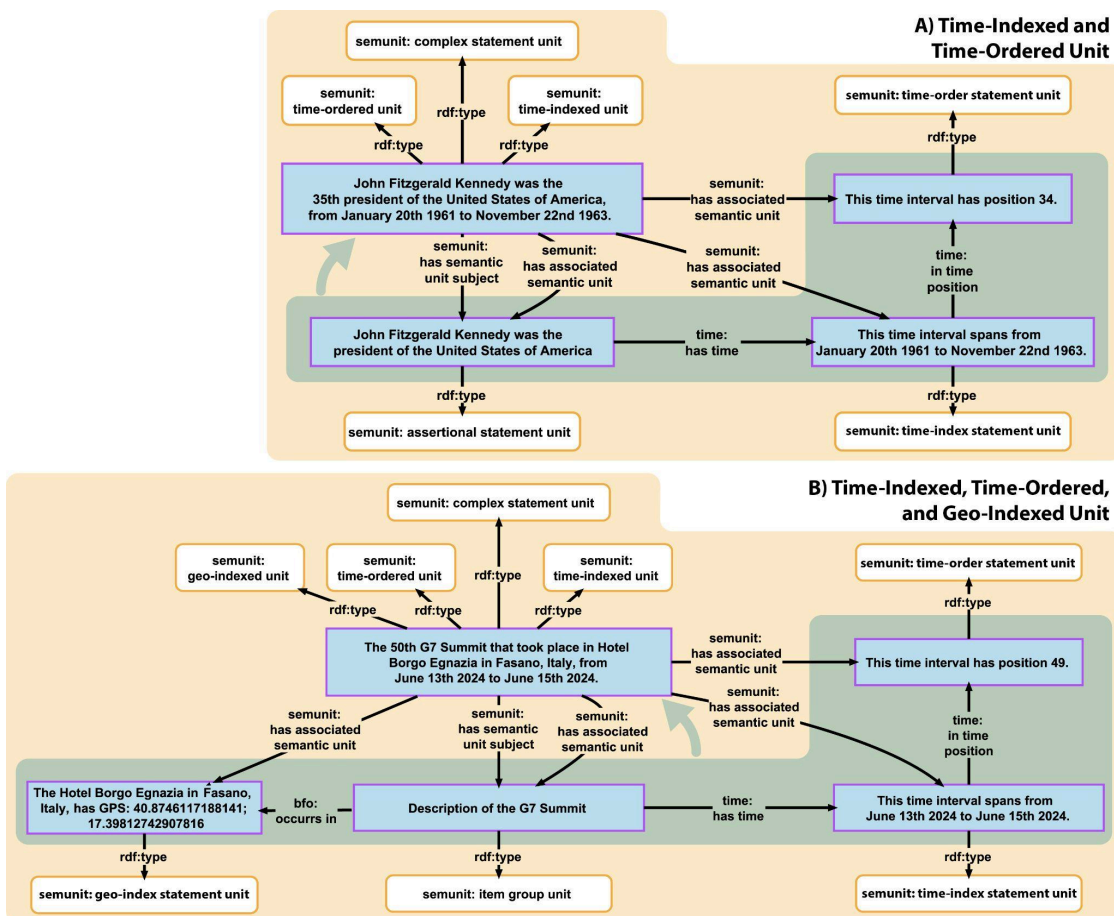

**Figure 17: Modelling of time-ordered, time-indexed, and geo-indexed statements using time-ordered, time-indexed, and geo-indexed units**. **A)** Modelling the statements from *Challenge 9* about the presidency of John F. Kennedy as a semantic unit. **B)** Modelling of a G7 Summit with a specification of its time-interval, its position within the sequence of G7 Summits, and its location as a semantic unit. *For reasons of clarity, metadata for each semantic unit is not represented.*

# Representing directive statements in FAIR knowledge graphs

**Challenge 10: Modelling directive statements**
**Directive statements** express prescriptive knowledge—**what ought to be done**—rather than descriptive knowledge about what *is*. They are typically realized by an actor executing a plan specified in a method or a list of instructions. Directive statements form the backbone of procedural, normative, and goal-oriented domains such as medicine, scientific **methods**, industrial **workflows**, engineering maintenance, or even simple everyday **tasks** like preparing food.

Consider the case of *boiling an egg*. This deceptively simple process embeds a series of directive components: the desired outcome (objective: a boiled egg), the necessary resources and their specifications (e.g., eggs, 5 liter water, cooking pot), necessary devices and their settings (e.g., stove and heat level), ordered action steps (e.g., "*heat water, then add egg*"), and sometimes conditional instructions (e.g., "*turn off the heat if the water boils*"). Directive statements may be explicit or implicit, and frequently omit the agent performing the action (e.g., "*Boil the egg!*" implies an unspecified actor).

From a formal semantics perspective, directive statements do not describe states of the world, but rather prescribe actions or intentions to achieve future states. Despite their centrality in task execution and instruction modelling, most mainstream knowledge representation formalisms—such as RDF, OWL, or labeled property graphs—do not natively support directive semantics. These systems are optimized for capturing **declarative (assertional) knowledge—things that are the case**—rather than imperative or **prescriptive knowledge—things that should be done**. Consequently, they lack constructs for representing goals, plans, obligations, or intentions, and offer no principled way to distinguish between directive variants.

To further complicate matters, directive statements may map onto different epistemic modalities, reflecting uncertainty, generality, or statistical likelihood. Consider the following forms:

1) **Assertional directive**: "*Make: Swan Anton is white!*".
2) **Contingent directive**: "*Make: Some swan is white!*".
3) **Prototypical directive**: "*Make: Most swans are white!*".
4) **Universal directive**: "*Make: Every swan is white!*".

Each of these expresses a different degree of generalization or strength of expectation, yet all carry prescriptive intent. Current ontology languages cannot accommodate this spectrum of directive forms, leaving a substantial modelling gap in domains where procedural and normative knowledge is essential.

**Approach: Directive unit**
To address this gap, we introduce the concept of the **directive unit**, a semantic unit designed to encapsulate directive statements. Analogous to question units introduced in *Challenge 6*, directive units provide a standardized wrapper for statements expressing what *should* be the case, according to a specific plan, protocol, or objective.

Directive units are defined by classifying a statement unit additionally as an instance of a *SEMUNIT: directive unit* (Fig. 18). Based on the underlying epistemic modality of the source statement, the directive unit may be further classified into one of four subtypes: assertional directive, contingent directive, prototypical directive, or universal directive unit.

This classification preserves the granularity of directive semantics and allows fine-grained reasoning about intent, obligation, and generality in prescriptive knowledge.

For example, the assertional directive "*Make: Swan Anton is white!*" is represented as an assertional directive unit, while "*Make: Most swans are white*" becomes a prototypical directive unit. Each unit preserves the original statement structure while encoding its imperative nature.

The directive unit model thus establishes a formal, interoperable structure for embedding prescriptive semantics in knowledge graphs, supporting use cases across domains where processes, workflows, and normative goals must be represented alongside facts.

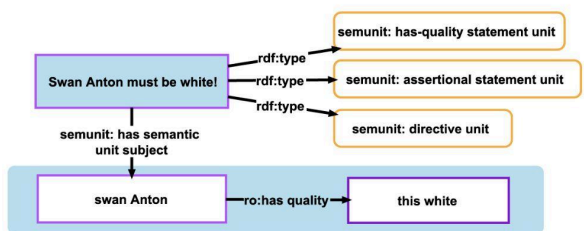
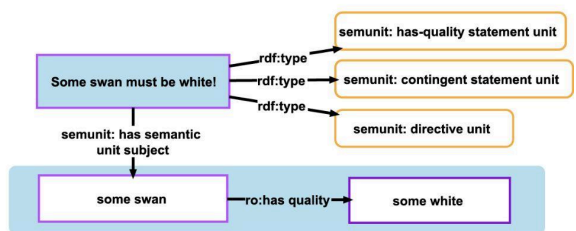
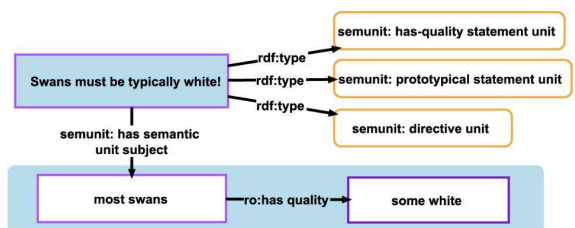
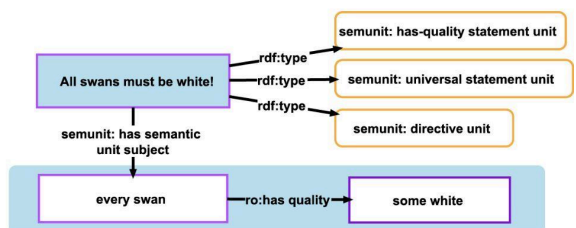

**Figure 18: Directive units**. Modelling the four directive statements from *Challenge 10* as directive units. *For reasons of clarity, metadata for each semantic unit is not represented.*

# Representing directive conditional statements in FAIR knowledge graphs

### Challenge 11: Modelling directive conditional statements

**Directive conditional statements** represent a fundamental and complex category of knowledge frequently encountered in procedural, scientific, and instructional contexts. These statements articulate **what must be done, given that a certain condition holds true**. They structurally combine two elements: an assertional statement in the condition (the *if-clause*) and a directive statement in the consequence (the *then-clause*).

For instance, consider the instruction: "*Add 2 kg of potatoes to 10 liters of water **if** the water has a temperature of 90°C!*". This statement binds the assertional statement *"the water has a temperature of 90°C"* to the directive statement *"add 2 kg of potatoes to the water"*. Such constructs are ubiquitous in domains such as medicine ("*Administer 500 mg of drug X if systolic blood pressure exceeds 160 mmHg*"), engineering ("*Open valve B if pressure in tank A exceeds threshold*"), and everyday procedural contexts like recipes and laboratory protocols.

Directive conditional statements are a special subclass of conditional statements, distinguished by their prescriptive *then-clause*. Not all conditional statements are directive; some are purely logical, analytical, or descriptive. For example, the philosophical proposition: "*There is no X that is good AND not beautiful*", can be logically reformulated as conditional statement: "*If X is good, then X is beautiful*" (72), which, in turn, may be expressed as universal statement: "*Everything that is good is also beautiful*" (72). Such transformations illustrate how conditional logic supports reasoning across both descriptive and prescriptive knowledge spaces.

Conditional statements, in general, link two or more propositions—an **if-clause** and a **then-clause**, sometimes including **else-clauses**. These proposition may themselves be compound, involving Boolean operators (AND, OR, XOR, NOT, EQUAL), increasing their complexity.

Modelling conditionality—especially when involving *statements about statements*—poses a deep challenge for most semantic technologies. RDF and OWL, as well as property graph systems like Neo4j, lack robust native support for higher-order constructs and complex reification mechanisms. Encoding a conditional relationship between two full statements (as opposed to linking two entities via a simple predicate) quickly becomes unwieldy using blank nodes, named graphs, or singleton properties. As a result, despite their prevalence, directive conditionals remain largely unmodelled in mainstream knowledge graph systems, leaving this rich and critical class of information outside the domain of current semantic systems, limiting their applicability to instruction-heavy domains.

**Approach: Directive conditional unit**

To address this challenge, we propose modelling directive conditional statements using a directive conditional unit, a structured semantic unit that explicitly relates a condition (assertional) to an action (directive). This semantic unit type builds upon the broader notion of a conditional unit, which represents the logical link between two propositions.

Conditional unit

At its core, a conditional unit is defined as a semantic relationship between two statements:

- an **if-clause** that is always an assertional statement unit,
- and a **then-clause** that can be either a directive or a logical conclusion.

In more advanced cases, either the *if-* or *then-clause* may involve compound logical expressions, such as: "*If temperature is >90°C AND pressure is <1 atm, then vent the chamber.*" To model such complexity, Boolean operators (AND, OR, NOT, XOR, EQUAL) are handled via **Boolean units** (see above), which allow multiple statement units to be composed into a single propositional block. This Boolean structure can serve as the input for either clause of a conditional unit, further extending the expressivity and realism of the model.

Directive conditional unit

When the then-clause is a directive unit, the resulting structure is a directive conditional unit (Fig. 19). This specific type of semantic unit supports a compositional logic where both parts of the conditional are independently modeled and then semantically linked. The directive conditional unit thus becomes a **complex statement unit**, one that encodes not only the truth-conditional semantics of its parts, but also their procedural interdependency.

Technically, the directive conditional unit is constructed as follows:

- The **if-clause** (an assertional statement unit) is linked to the **then-clause** (a directive unit) via the property *SEMUNIT:is if of then*.
- Both the *if* and *then* statements are associated with the overarching **conditional unit resource** using *SEMUNIT:has associated semantic unit*.
- The system can dynamically generate human-readable labels by combining the **dynamic labels** of the assertional statement unit and the directive unit, adding in front of the former "*If*" and of the latter "*then*".

This formalization enables knowledge graphs to represent not just facts or static entities, but dynamic, contingent procedures that are foundational to action-based reasoning, automated execution, and domain-specific instruction systems.

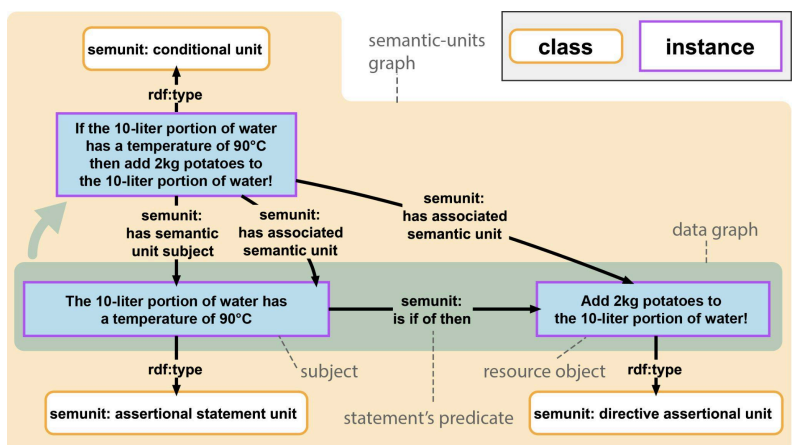

**Figure 19: Directive conditional unit**. Modelling of an if-then directive conditional statement as a directive conditional unit consisting of an assertional statement unit as the if-clause and a directive assertional statement unit as the then-clause. *For reasons of clarity, metadata for each semantic unit is not represented.*

# Representing logical arguments in FAIR knowledge graphs

**Challenge 12: Modelling logical arguments**

Logical arguments represent a core structure in human reasoning, particularly within scientific discourse. They enable us to **justify claims**, **generate hypotheses**, and **draw conclusions from evidence**. Formally, a logical argument links two premises to a conclusion by way of a specific mode of reasoning—deductive, inductive, or abductive. Each of these argument types consists of three semantic components:

- a **rule-clause** (a general statement or principle),
- a **case-clause** (a particular fact or observation),
- and a **result-clause** (a deduced, induced, or inferred consequence).

Depending on the direction and modality of reasoning, the arrangement of these clauses shifts, resulting in three canonical patterns (73):

> **Deductive argument**
> premise (rule):        "*All swans are white.*"           universal statement
> premise (case):       "*X is a swan.*"                    assertional statement
> conclusion (result):  "*X is white.*"                    assertional statement
> The conclusion **necessarily** follows.

**Inductive argument**

| | | |
|---|---|---|
| premise (case): | *"X is a swan."* | assertional statement |
| premise (result): | *"X is white."* | assertional statement |
| conclusion (rule): | *"All swans are white."* | universal statement |

The conclusion is **probable** (verisimilar).

**Abductive argument**

| | | |
|---|---|---|
| premise (result): | *"X is white."* | assertional statement |
| premise (rule): | *"All swans are white."* | universal statement |
| conclusion (case): | *"X is a swan."* | assertional statement |

The conclusion is **possible** (plausible hypothesis).

All three forms are logically structured conditionals. The premises jointly form the **if-clause**, while the conclusion serves as the **then-clause** (see *Challenge 11*). However, the **modality** of the conclusion varies:

1) *Then-necessarily* in deduction,
2) *Then-probably* in induction,
3) *Then-possibly* in abduction.

Despite their foundational role in science and logic, **logical arguments remain poorly supported** in mainstream semantic technologies. RDF/OWL and property graph models (e.g., Neo4j) lack constructs for natively expressing **multi-premise arguments**, let alone representing their underlying reasoning modality or inferential structure—they already lack constructs for representing universal and directive conditional statements in a knowledge graph (see *Challenges 2 and 11*)). These platforms also struggle with compound clauses, reification of logical dependencies, and distinctions in conclusion boldness. Consequently, even simple arguments are represented outside the knowledge graph and are often relegated to external rule engines or remain informal, undermining the utility of knowledge graphs in scientific and analytical contexts.

**Approach: Logical argument units**

To overcome these limitations, we introduce the concept of a **logical argument unit**—a specialized conditional unit that encapsulates structured reasoning. Each logical argument unit links **three distinct statement units** (73):

- A **case statement** (assertional statement or named-individual identification unit),
- A **rule statement** (universal, prototypical, or contingent statement unit),
- A **result statement** (assertional statement or named-individual identification unit).

These statement units serve as either **premises** or the **conclusion**, depending on the type of logical argument being modeled. Thus, each argument type corresponds to a unique configuration of clause roles and inferential modality.

### Types of logical argument units

A knowledge graph can be used to formally represent the three different types of logical arguments by organizing their premise-clauses and conclusion-clauses into three corresponding types of **logical argument units** (see Fig. 20):

1. **Deduction Unit**
   Case + Rule ⇒ Result
   (Conclusion: assertional; modality: necessary)
2. **Induction Unit**
   Case + Result ⇒ Rule
   (Conclusion: universal/prototypical/contingent; modality: probable)
3. **Abduction Unit**
   Result + Rule ⇒ Case
   (Conclusion: assertional; modality: possible)

This structure builds upon the broader semantic unit framework introduced in previous challenges. Logical argument units are modeled using subproperties of *SEMUNIT:has associated semantic unit* (Fig. 20). These enable precise associations of **premise-clauses** and **conclusion-clauses** with the respective logical argument unit. Where clauses are composite (e.g., "*X is white* AND *X has wings*"), Boolean units are used to join multiple statement units.

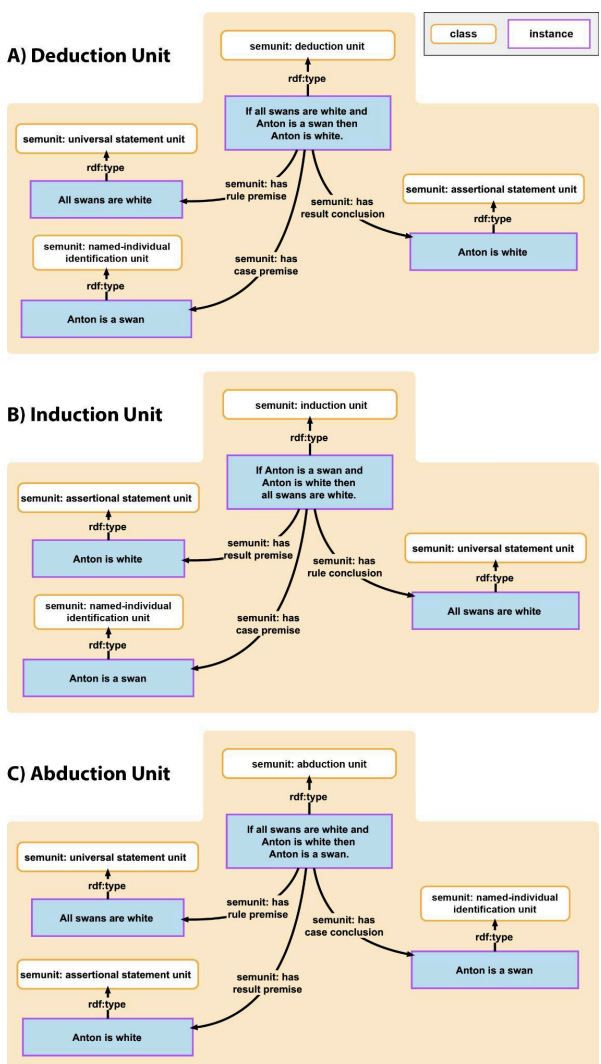

**Figure 20: Logical argument units**. Modelling of logical arguments as **A)** a deduction unit, **B)** an induction unit, and **C)** an abduction unit. *For reasons of clarity, metadata for each semantic unit is not represented.*

Modelling boldness in inductive conclusions

Inductive reasoning introduces a critical nuance: **conclusion boldness**. The inferred rule may differ in generality and evidential burden:

- **Universal**: "*All swans are white*" (high boldness),
- **Prototypical**: "*Most swans are white*" (moderate boldness),
- **Contingent**: "*Swans can be white*" (low boldness).

While a single observation (e.g., "*Swan Anton is white*") suffices to justify the contingent claim ("*Swans can be white*"; a truth-preserving conclusion), it is insufficient for prototypical or universal claims (a not truth-preserving conclusion). The latter require additional empirical support: demonstrating, for instance, that white swans outnumber non-white swans (for prototypical) or that no non-white swans exist (for universal). Thus, boldness reflects the **epistemic cost of justification**, and modeling it is essential for semantically precise scientific reasoning.

## Hypothetical versus factual conclusions in abduction

Abductive conclusions are **plausible, but not certain**. For example, from "*Anton is white*" and "*All swans are white*," one may infer "*Anton is a swan.*" This may be modeled:

- As a **hypothesis** (e.g., "*Anton could be a swan*"),
- Or as a **presumed fact** (e.g., "*Anton is a swan*").

The modeling framework supports both interpretations, with the hypothesis form reflecting lower boldness and more conservative inference.

## Logical reasoning in practice

Logical argument units enable the specification of **inference rules** within the graph. These rules can be leveraged in reasoning systems to derive new conclusions, mirroring deductive, inductive, or abductive logic. For instance:

- If both premises in a deduction unit are present, a reasoner can assert the conclusion.
- Inductive reasoning rules can vary based on the boldness of the inferred generalization.
- Abductive rules may produce tentative hypotheses that can be validated through further observation or cross-referenced data.

These reasoning mechanisms complement existing OWL reasoning but extend well beyond it, supporting **multi-premise, modality-sensitive inference** grounded in formal semantic structures.

The introduction of **logical argument units** marks a pivotal step toward empowering knowledge graphs with true logical expressivity. By capturing the structure and modality of deductive, inductive, and abductive reasoning, the framework transcends static data representation and moves toward **semantic reasoning as a first-class capability**. It also provides the foundational scaffolding for integrating argumentative structures into knowledge-centric AI, scientific publishing, hypothesis tracking, and intelligent agents.

When combined with the modeling of universal statements, directive conditionals, and Boolean units (as addressed in previous challenges), this approach delivers a coherent, extensible paradigm for semantically rich and logically grounded knowledge representation.

# Discussion

## Taxonomy of semantic unit types

Over the course of the previous sections, we introduced a comprehensive set of semantic unit types to address a range of representational challenges in knowledge graphs. These semantic units serve as the foundational building blocks for categorizing and structuring knowledge, encompassing a spectrum of information types—from assertional and universal statements to conditional and logical argument units.

To provide an overarching view, Figure 21 presents the taxonomy of all semantic unit types developed in this work. This taxonomy organizes the units hierarchically, distinguishing between:

- **Statement units** (e.g., *named-individual identification units*, *assertional statement units*),
- **Complex statement units** as a subclass of statement units (e.g., *conditional units*, *directive units*, *epistemic units*),
- **Compound units** used for designating semantically meaningful collections of statement and complex statement units (e.g., *item units, class profile units, granularity tree units*),
- and **additional semantic units** that serve the purpose to classify their content (e.g., *Boolean units*, *negation units, cardinality restriction units, question units*).

Each semantic unit class corresponds to a specific kind of information structure with a defined internal composition and external linking mechanism. Importantly, this taxonomy also makes explicit the **logical capabilities** of each unit type—information that plays a critical role in determining their compatibility with formal reasoning frameworks (see next section).

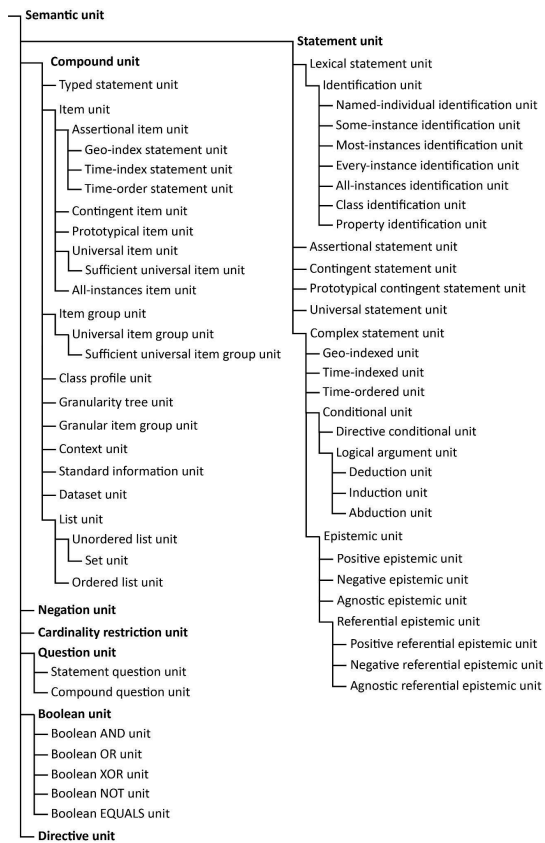

**Figure 21: Classification of different categories of semantic units.**

# Semantic modularization, unit granularity, and logical reasoning in OWL-extended knowledge graphs

One of the central challenges in extending OWL-based knowledge graphs is the **need to represent knowledge that exceeds OWL's formal expressivity**, without undermining the graph's overall logical coherence. Standard OWL reasoners are strictly confined to Description Logics-based semantics, and any departure from this—such as the inclusion of probabilistic, defeasible, procedural, or abductive knowledge—introduces ambiguity or renders reasoning infeasible. There are, broadly speaking, **two unsatisfactory alternatives** to addressing this gap:

1. **Avoid modeling such information altogether**, leaving it outside the knowledge graph as informal metadata or external documentation.
2. **Include non-OWL content within the knowledge graph**, but without distinguishing it from OWL-valid content—thereby obscuring what is logically valid, what is uncertain, and what is merely descriptive.

The **semantic unit approach** provides a structured alternative to both. Since each piece of information is encapsulated within a *typed* semantic unit, it becomes possible to **explicitly distinguish which parts of the graph are reasoning-capable and under which logical framework**, providing a **context-aware formalism**. For example:

- An assertional statement unit is fully compatible with OWL Description Logics reasoning.
- An **induction unit** producing prototypical statements may require probabilistic reasoning (e.g., Bayesian or statistical frameworks).
- A **directive unit** may encode procedural rules not formally compatible with monotonic reasoning at all.

This makes it feasible to organize the knowledge graph **modularly**, with clear boundaries between logically interpretable and non-interpretable sections. This modular organization constitutes a form of **semantic modularization**—the deliberate decomposition of the knowledge graph into **logic-aware**, semantically meaningful units that can be independently annotated, reasoned over, or ignored, depending on the task at hand. Each semantic unit class that is associated with a specific set of frameworks and strategies for their formal semantics can be extended with additional properties to indicate:

- The **logical framework** applicable to each semantic unit instance (e.g., OWL Description Logics, rule-based logic, probabilistic logic, abductive inference),
- The **associated reasoner or engine** capable of processing it (e.g., HermiT, Pellet, custom SPARQL engine, external ML-based inference).

This architectural separation offers several key benefits:

- **Transparent reasoning scope**: Consumers of the graph know exactly which inferences are logically grounded and which are not.
- **Mixed reasoning support**: The graph can simultaneously support multiple reasoning paradigms without conflation or inconsistency.

- **Safe extensibility**: Future logic types (e.g., temporal, default, or legal reasoning) can be integrated without redesigning the existing data and knowledge.

From both a design and usability perspective, **embedding logic-type annotations at the semantic unit level** presents a powerful mechanism for organizing and managing complex knowledge systems. It allows knowledge graphs to evolve into **multi-logic, heterogeneous reasoning environments**, where each part of the graph is epistemically transparent.

Rather than attempting to force all knowledge into a single logical framework—or worse, discarding formal reasoning entirely or data and knowledge that cannot be expressed within the constraints of Description Logics—the semantic unit architecture enables **selective formalism**, where reasoning is context-sensitive and semantically traceable. This makes the approach not only theoretically sound, but also highly adaptable to real-world data modeling and semantic AI applications.

By **embedding logical transparency** at the level of individual semantic units, **semantic modularization enables a flexible reasoning environment** where inference scope is explicitly defined and epistemic boundaries are respected.

## On the generality of the semantic unit approach

While the concept of **semantic units** has been operationalized in this work within the frameworks of RDF/OWL-based knowledge graphs and property graphs such as Neo4j (27), we emphasize that the underlying principle is **technology-agnostic**. At its core, the semantic unit approach is based on the idea that **statements are the smallest semantically meaningful units of information**—they should be **first-class citizens** in any knowledge management system, atomic constituents of human communication about data, facts, and knowledge. As such, the approach should not be limited to knowledge graphs, but rather seen as a **general paradigm for organizing information across heterogeneous data systems**.

The central insight is this: for any dataset to be human-interpretable and communicable, its content must be expressible—at least in principle—in the form of one or more natural language statements. Consequently, structuring data into **statement units** represents a format-independent abstraction for information management, one that prioritizes **semantic legibility** and **compositional modularity**.

To generalize the semantic unit framework across technologies and data formats, a dataset must satisfy the following **five minimal structural requirements**:

1. **Partitionability into statement units**: The dataset must be divisible into non-overlapping subsets, each corresponding to a semantically meaningful statement interpretable by a human. Each relation between two or more atomic data points must belong to one and only one such subset.
2. **GUPRI Assignment**: Each such statement unit must be assigned its own **GUPRI**, which functions as both the identifier for the statement unit and the pointer to its underlying data subset.
3. **Composable Referencing**: It must be possible to reference one semantic unit from another via their GUPRIs—**without duplicating content**—so that compound units can be constructed from collections of other semantic units.

4. **Retrievability**: The full data subset represented by a semantic unit must be retrievable using its GUPRI, either directly or through composition.
5. **Typed Instantiation**: Each semantic unit must instantiate a formally defined class or type (e.g., assertional statement, conditional unit, logical argument unit) that declares its semantic structure and interpretive expectations.

These conditions are already conceptually supported by the **FAIR Digital Object (FDO)** framework (74), which aims to make all digital artifacts FAIR. FDOs offer a format-neutral, identifier-centric abstraction layer, which—when aligned with the semantic unit paradigm—enables **technology-independent implementation** of semantically structured data systems.

By adopting the semantic unit model within the FDO ecosystem (see also discussion in (25)), statement-based modularity can be extended to non-graph-based environments, including document repositories, tabular data, relational databases, and file-based archives. Thus, semantic units can serve as a **unifying logic for data modeling**, regardless of the underlying format or infrastructure.

# Conclusion

With semantic units, we introduce a principled method for structuring knowledge graphs into semantically meaningful and identifiable subgraphs, each equipped with its own GUPRI, implementing **semantic modularization** by organizing the knowledge graph into logic-aware, cognitively accessible modules that can be independently annotated, queried, or reasoned over. This layered approach complements the introduction of some-instance, most-instances, every-instance, and all-instances resources—new types of **representational entities** (75) beyond conventional instances, classes, and properties. Together, these elements significantly enhance the **expressivity**, **granularity**, and **cognitive interoperability** of knowledge graphs.

Semantic units function as **higher-order representational entities**, standing at a level of abstraction above traditional RDF triples. While technically modeled as instances of defined ontology classes, semantic unit resources semantically embody **complete statement graphs**—i.e., individual statements or coherent clusters of related statements—allowing domain knowledge to be packaged, referenced, and reasoned over in modular form. In doing so, semantic units bridge the gap between granular data representation and human-level semantic understanding.

By distinguishing between assertional, contingent, prototypical, universal, and lexical statement units as top-level categories of statement units, and by incorporating alternative instance quantifiers (some, most, every, all), the semantic unit framework provides a **formal scaffolding** for representing both factual knowledge and **contextual, epistemic, or generalizing statements**. This approach allows us to:

- Address *Challenge 1* by enabling formal semantics for contingent and prototypical statements;
- Resolve *Challenge 2* by expressing universal statements as ABox constructs, thereby including them in the domain of discourse;
- Overcome *Challenge 3* by representing triangular class axioms in a way that avoids OWL's known conceptual constraints;
- Tackle *Challenge 4* with the introduction of negation and cardinality restriction units as ABox-level entities;

- Address *Challenge 5* by representing epistemic perspectives and disagreements through complex statement units.

Each statement unit is self-contained: it references a clearly defined **data schema** (e.g., SHACL shape) in its metadata, is typed via a corresponding **statement unit class**, and can be linked to a curated set of **CRUD query patterns**. This design ensures that the contents of each semantic unit are **FAIR**, particularly with respect to **schema interoperability** (8) (Challenge 6). Furthermore, because every triple in the knowledge graph is mapped to exactly one statement unit, the entire graph becomes **structurally transparent and semantically annotated**.

The benefits extend to **user interaction** and **knowledge accessibility**. By enabling **question units** to use named-individual, some-instance, most-instances, and every-instance resources in subject and object positions of statement units, the framework supports natural language-like querying (*Challenge 7*). By utilizing the underlying data schemata, scripts can be developed that transform question units into readily executable SPARQL or Cypher queries, allowing users to specify queries in the form of question units without requiring any knowledge of graph query languages.

Semantic units also support the creation of **standard information units**—predefined, domain-specific views that encapsulate commonly referenced or widely accepted knowledge fragments (*Challenge 8*). This feature fosters **reusability and citation** of knowledge components across datasets.

With the introduction of complex statement units, modelling (ordered series of) time-indexed statements about (consecutive) events or time- and location-dependent states in a knowledge graph becomes straightforward (*Challenge 9*) as well as the specification of conditional if-then statements. Based on the distinction of assertional, contingent, prototypical, and universal statements, we can also differentiate between the corresponding forms of directive statements in a knowledge graph using directive units (*Challenge 10*) and we can combine the representational formalism for conditional if-then statements and directive statements to model directive conditional statements in a knowledge graph using directive conditional units (*Challenge 11*).

Finally, we have demonstrated how semantic units can be used for modelling logical arguments as logical argument units in the form of deduction, induction, and abduction units and have briefly discussed how they can be applied for reasoning within a knowledge graph (*Challenge 12*).

From both theoretical and practical standpoints, the semantic unit approach presents a **general, extensible framework** for organizing digital information around human-meaningful statements. By decoupling from the constraints of specific graph formalisms (e.g., RDF/OWL or property graphs) and aligning with broader infrastructures like the **FDO** ecosystem, the model supports **cross-format, cross-platform implementation**.

This abstraction does not merely facilitate better knowledge graph engineering—it proposes a **universal architecture for semantic decomposition and modular knowledge representation**. In doing so, it establishes semantic units as a foundational principle for knowledge representation, one that spans technologies and supports a more modular, interpretable, and communicable data ecosystem.

The enduring value of the semantic unit framework is not limited to its enhanced expressivity for automated reasoning—it stems from a more fundamental insight: **not all knowledge that matters to humans can—or should—be fully captured within the bounds of machine reasoning.** Formal semantics inevitably reach their limits when confronted with contingent, epistemic, or generalizing knowledge. Yet these kinds of statements are pervasive in science and everyday reasoning.

Rather than discarding such knowledge or misrepresenting it through forced formalization, semantic units offer a way to document, index, and interlink these statements in a structured and findable way by **semantic modularization**. Even when certain claims cannot be interpreted by OWL reasoners, they can still be **addressed, understood, and acted upon** by human agents and machine-assisted systems. This dual capability is vital for real-world knowledge infrastructures.

In this light, semantic units are more than a technical innovation: **they constitute a representational philosophy**. They allow knowledge graphs to speak in ways that OWL alone cannot—preserving meaning, traceability, and contextual richness. With them, *OWLs no longer have to fully understand everything they say*—but they can still say it clearly, consistently, and FAIRly. **Structuring knowledge for human understanding is as critical as enabling machine reasoning**. In this light, *semantic units are not just a technical construct—they are a bridge between human semantics and machine syntax*. **Semantic modularization through semantic units** offers a scalable and extensible foundation for constructing FAIR-aligned knowledge graphs that can integrate heterogeneous logical frameworks while remaining comprehensible and reusable.

# Abbreviations

| | |
|---|---|
| ABox | Assertion Box |
| ARDS | Acute Respiratory Distress Syndrome |
| BFO | Basic Formal Ontology |
| CHEBI | Chemical Entities of Biological Interest ontology |
| CSV | Comma-Separated Values |
| FAIR | Findable, Accessible, Interoperable, and Reusable |
| FOAF | Friend of a Friend ontology |
| FOODON | Food Ontology |
| IAO | Information Artifact Ontology |
| LinkML | Linked Data Modeling Language |
| LLM | Large Language Model |
| NCBITaxon | National Center for Biotechnology Information Taxon Ontology |
| NCIT | National Cancer Institute |
| OBI | Ontology for Biomedical Investigations |
| OWA | Open World Assumption |
| OWL | Web Ontology Language |
| PATO | Phenotype and Trait Ontology |
| PO | Plant Ontology |
| RDF | Resource Description Framework |
| RDFS | RDF-Schema |
| RO | OBO Relations Ontology |
| SEMUNIT | Semantic Unit Ontology |
| SHACL | Shape Constraint Language |
| SPARQL | SPARQL Protocol and RDF Query Language |
| TBox | Terminology Box |
| UBERON | Uber-anatomy ontology |
| UI | User Interface |
| UO | Units of Measurement Ontology |
| GUPRI | Globally Unique Persistent and Resolvable Identifier |
| URI | Uniform Resource Identifier |
| XSD | Extensible Markup Language Schema Definition |

# Declarations

## Ethics approval and consent to participate

Not applicable

## Consent for publication

Not applicable

## Availability of data and materials

Not applicable

## Competing interests

The authors declare that they have no competing interests

## Funding

Lars Vogt received funding by the ERC H2020 Project 'ScienceGraph' (819536) and by the project DIGIT RUBBER, Grant no. 13XP5126B (TIB), and InSuKa, Grant no. 13XP5196F, both funded by the German Federal Ministry of Education and Research (BMBF).

## Author's contributions

LV: developed the concept of semantic units and wrote the initial manuscript. RH: developed the formal semantics described in *Logical semantics of semantic units*, and he wrote this chapter. All authors: reviewed the manuscript.

## Acknowledgements

We thank Werner Ceusters, Nico Matentzoglu, Manuel Prinz, Marcel Konrad, Philip Strömert, Roman Baum, Björn Quast, Peter Grobe, István Míko, Manfred Jeusfeld, Manolis Koubarakis, Javad Chamanara, and Kheir Eddine for discussing some of the presented ideas. We are solely responsible for all the arguments and statements in this paper.